\documentclass[twocolumn]{revtex4}

\usepackage{graphicx}

\def\bea{\begin{eqnarray}}
\def\eea{\end{eqnarray}}
\def\be{\begin{equation}}
\def\ee{\end{equation}}
\def\fr{\frac}
\def\ci{\cite}
\def\vp{\varphi}
\def\ga{\gamma}

\def\le{\left}
\def\ri{\right}

\begin{document}

\title{Quantum Generation of Dark Energy}
\author{A. de la Macorra and F. Briscese }
\affiliation{Instituto de F\'{\i}sica, Universidad Nacional
Autonoma de Mexico, Apdo. Postal 20-364,
01000 M\'exico D.F., M\'exico\\
Part of the Collaboration Instituto Avanzado de Cosmologia}

\begin{abstract}
\begin{center}
\end{center}

We present a type of dark energy models where  the particles of
dark energy $\phi$  are dynamically produced via a quantum
transition at very low energies. The scale where the transition
takes places depends on the strength $g$ of the interaction
between $\phi$ and a relativistic field $\vp$. We show that a
$g\simeq 10^{-12}$ gives a generation scale $E_{gen} \simeq 1\,eV$
with a cross section $\sigma\simeq 1 \,pb$ close to the WIMPs
cross section $\sigma_{w}\simeq pb$ at decoupling.  The number
density $n_\phi$  of the $\phi$ particles is a source term in
the  equation of  motion of $\phi$ that generates the scalar
potential $v(\phi)$ responsible for the late time acceleration of
our universe. Since the appearance of $\phi$ may be at very low
scales, close to present time, the cosmological coincidence
problem can be explained simply due to the size of the coupling
constant. In this context it is natural to unify dark energy with
inflation in terms of  a single scalar field $\phi$. We use the
same potential $v(\phi)$ for inflation and dark energy. However,
after inflation $\phi$ decays completely and reheats the universe
at a scale $E_{RH} \propto h^2 \, m_{Pl}$, where $h$ is the
coupling between the SM particles and $\vp$. The field $\phi$
disappears from the spectrum during  most of the time, from
reheating until its  re-generation at late times, and therefore it
does not interfere with the standard decelerating radiation/matter
cosmological model allowing for a successful unification scheme.
We show that the same interaction term that gives rise to the
inflaton decay accounts for the late time re-generation of the
$\phi$ field giving rise to dark energy. We present a simple model
where the strength of the $g$ and  $h$ couplings are set by the
inflation scale $E_I$ with $g=h^2 \propto E_I/m_{Pl}$ giving a
reheating scale $E_{RH} \propto E_I $ and $\phi$-generation scale
$E_{gen} \propto E_I^2/m_{pl} \ll E_{RH}$. With this
identification we reduce the number of parameters and the
appearance of dark energy is then given  in terms of the inflation
scale $E_I$.

\end{abstract}

\maketitle

\section{Introduction.}

The nature and dynamics of Dark Energy ''DE'', which gives the
accelerating expansion  of the universe at present time, is now
days one of the most interesting and stimulating fields of
physics. It was discovered more than ten years ago \cite{first
aceleration} and it has been confirmed by further cosmological
observations, being now one of the most robust conjecture  in
modern physics. In fact the acceleration of the present universe
has many experimental proofs such as  the CMB temperature and
fluctuations \cite{komatsu wmap}, in the matter power spectrum
measured by galaxy surveys \cite{sdss,Tegmark SDSS} and in type Ia
supoernovae \cite{Riess IA supernova,Kowalski supernova,Miknaitis
supernova}.

The most popular model is the so called $\Lambda$CDM model, in
which a cosmological constant and some amount of cold dark matter
are included ''by hand''. Despite  its extraordinary consistence
with observations, $\Lambda$CDM  is an effective model that leaves
many unsolved theoretical question. In fact the existence of the
cosmological constant and its order of magnitude have   no
theoretical justification in  $\Lambda$CDM. The cosmological
coincidence problem, that is why the universe is starting to
accelerate right now, is also unsolved. Introducing a cosmological
constant at the initial stages of the standard cosmological model
is specially troublesome since one has to fine tune its value to
one parte in $10^{120}$. This problem can be ameliorated if we
understand when and how dark energy appears in the universe and
this is the main motivation of our present work \cite{axelfab}.
However, our approach will also help us to   unify dark energy
with inflation.

An attractive dark energy alternative to the $\Lambda$CDM model
consists in introducing a ''quintessence'' scalar field $\phi$
that generates the accelerating expansion \ci{Q}\ci{axQ} of the
universe due to is dynamics. The dynamics is fixed by its
potential $v(\phi)$ and it is possible to choose potentials that
lead to a late time acceleration of the universe \ci{axQ}. This
scalar field can be a fundamental  or composite particle as for
example bound states \ci{axDG}. In the second case, the bound
quintessence fields are scalar fields composed of fundamental
fermions, such as meson fields, and  can be generated at low
energies as a consequence of a low phase transition scale due to a
strong gauge coupling constant \ci{axDG}. This allows to
understand why DE appears at such late times. On the other hand,
in the former case the appearance of fundamental scalar field is
right at the beginning of the reheated universe and the
acceleration of the universe takes place at a much later time due
to the classical evolution of the quintessence field $\phi$. The
huge difference in scales between the reheating and dark energy
scales requires a fine tuning in the choice of the potential.

Here, we present an interesting alternative, namely that the
emergence of the fundamental quintessence particles $\phi$ is
originated from a quantum transition taking place at low energies,
e.g. as low as $eV$  \cite{axelfab}. The scale where this
transition takes places depends on the strength of the interaction
between $\phi$ and a relativistic field $\vp$ and it is
dynamically determined by the ratio  $\Gamma/H$ where $\Gamma$ is
the transition rate   and   $H$ the Hubble constant. A
 value of the coupling $g\simeq 10^{-12}$ gives a generation scale
$E_{gen}\simeq 1\,eV$ with a cross section $\sigma = g^2/32 \pi
E_{gen} \simeq 1 \, pb$ close to the WIMPs cross section
$\sigma_{w}\simeq O(pb)$ at decoupling\ci{wimp}. The subsequent
acceleration of the universe is due to the classical evolution of
$\phi$ due to the scalar potential $v(\phi)$. Our quantum
generation scheme does not aim to derive the potential $v(\phi)$
but to understand why dark energy dominates at such a late time.
Clearly, by closing the gap between the energy today $E_o$, where
the subscript $o$ always refers to present time quantities,  and
that of $\phi$ production $E_{gen}$, we do not require a fine
tuning of the parameters in $v(\phi)$. Since the appearance of
$\phi$ may be at such low scales this offers a new interpretation
and solution to the cosmological coincidence problem in terms of
the size of the coupling constant $g$.

Furthermore, this late time production  of the $\phi$ particles
allows to implement in a natural way a dark energy-inflaton
unification scheme. In this scenario, after inflation the field
$\phi$ decays completely and reheats the universe with standard
model particles. The universe expands then in a decelerating  way
dominated first by radiation and later by matter. At low energies
the same interaction term that gives rise to the inflaton decay
accounts for the quantum re-generation of the $\phi$ field giving
rise to dark energy. In general, it is not complicated to choose a
scalar potential such that   the universe accelerates in two
different regions, at early inflation and dark energy epochs, as
in quintessential models \ci{scalarunification}. However, the
universe requires to be most of the time dominated first by
standard model relativistic particles  and later by matter. The
reheating of the universe and the long period of decelerating
phase are  usually not taken into account in inflation-dark energy
unification  models and these features are essential in the
standard cosmological Big Bang model. In our case, the
inflaton-dark energy field is completely absent during most of the
time (from reheating until re-generation) and therefore it does
not interfere with the standard cosmological model. We will
exemplify our inflation-dark energy unified scheme with a simple
model. The scalar potential $v(\phi)$ will have only two
parameters fixed by the conditions to give the correct density
perturbations $\delta \rho/\rho$ and the present time dark energy
scale. The two couplings $h,g$, which give the strength of
reheating process with SM particles and the $\phi$ re-generation
process at low energies, respectively, are free parameters but we
may take them as $g=h^2 \propto E_I/m_{pl}$, where $E_I$ is the
scale of inflation and it is one of the parameters of $v(\phi)$.
Therefore, starting with four free parameters we can reduce the
number to only two and these two are fixed by observations. This
gives a reheating scale $E_{RH} \propto E_I$ and $\phi$-generation
scale $E_{gen}\propto E_I^2/m_{pl} \ll E_{RH}$.

The paper is organized as follows: in section \ref{motivations} we
give an overview of the late time quantum generation of the
quintessence field $\phi$ and its possible unification with the
inflaton field. In section \ref{regeneration} we present the dark
energy quantum generation in detail. In section \ref{regeneration}
we show how to unify inflation and dark energy in terms of a
single scalar field in the context of our dark energy quantum
generation process and we present a simple model. Finally, in
section \ref{phenomenology} we present the main phenomenological
consequences of our model while in section \ref{conclusions} we
resume and conclude.

\section{Overview} \label{motivations}

To avoid any future confusion we state the terminology used in
this work. We define as usual the energy of the universe at any
given time as $E\equiv \rho^{1/4}$ or for a  $i$-species as
$E_i\equiv\rho_i^{1/4}$. For particles we take their energy $E_a$
as $E_a=(p_a^2+m_a^2)^{1/2}$, where  $p_a, m_a$ are the momentum
and mass of the $a$-particle. If the particles are relativistic
and in thermal equilibrium then one can define a temperature $T$
with an energy density  $\rho_a = \frac{\pi^2}{30} g_a T ^4$ and
number density $ n_a=\zeta(3)/\pi^2\,g_a\, T^3 $, with  $g_a$ the
relativistic degrees of freedom and $\zeta(3) \simeq 1.2$  is the
Riemann zeta function of $3$. The energy distribution of
thermalized particles is strongly piked around the mean value
$\bar E = \bar{r} \, T $  with $\bar{r}\equiv
(\rho/nT)=\pi^4/30\zeta(3) \simeq 2.7$ and we identify the energy
of the particle $E_a$ with the average energy $E_a=\bar E_a= \bar
r T$. We will loosely refer then to either the temperature or the
energy of the (thermalized) relativistic particles and
we will work in natural units with $m^2_{pl} \equiv 1/8\pi G =1$
but we will reintroduce the correct energy dimensions when
convenient.

We will first present the quantum generation process of dark
energy field $\phi$ and we will later discuss the possibility to
unify it with the inflaton field. As mentioned in the
introduction,  the main goal of this paper is to describe how the
dark energy field $\phi$ may be generated at a very late time via
a quantum transition process and how the scalar potential
$v(\phi)$, responsible for the late time acceleration, is
generated. Once $v(\phi)$ is produced, the classical equation of
motion gives the dynamics of $\phi$, and choosing an appropriated
flat $v(\phi)$ at low energies ensures that the universe enters an
accelerating epoch close to present time. This work does not aim
to derive the potential $v(\phi)$ but to understand why dark
energy dominates at such a late time. Of course, by closing the
gap between the energy today $E_o$  and that of $\phi$ production
$E_{gen}$ there is a no longer a fine tuning of the parameters in
$v(\phi)$.

We describe now the generation process. We take a universe filled
with standard model SM particles and an extra relativistic field
$\vp$ and no $\phi$ particles, i.e. $\Omega_\phi=0$. The $\vp$
particle is not necessarily contained in the SM (however, an
interesting possibility is to  associate $\vp$ with neutrinos) but
we require that at time of $\phi$-generation $\rho_\vp(E_{gen})>
\rho_{DE}(E_o)$. In the context of inflation-dark energy
unification the $\vp$ must have been  in thermal equilibrium with
the SM  and therefore $T_\vp\simeq T_\ga$ with $\Omega_\vp \simeq
(g_\vp/g_{rel}^{SM}) \Omega_{rel}^{SM}$, where $g_\vp$ and
$g_{rel}^{SM}$ are the relativistic degrees of freedom of $\vp$
and the SM respectively and $\Omega_{rel}^{SM}$ is the density of
SM relativistic particles. Since we want to produce the $\phi$
particles via a quantum transition process we couple it to $\vp$
via an interaction term $L_{int}$, e.g. $L_{int} = g \, \phi \,
\varphi^3$ or $L_{int} = g \, \phi^2 \, \varphi^2$. In order to
produce the $\phi$ particles at low energies we require that the
interaction rate $\Gamma$ of the quantum process must be initially
smaller than the Hubble parameter $H$ and the ratio $\Gamma/H$
should increases with the expansion  of the universe. This will
happen if $\Gamma$ decreases less rapidly than $H$. For example if
we have a $2\leftrightarrow 2$ process of relativistic particles
the transition rate scale as $\Gamma\propto g^2 T_\vp $ and
$H\propto T_\vp^2$ giving $ \Gamma/H\equiv T_{gen}/T_\vp \propto
g^2/ T_\vp$, where we have taken $T_\vp\simeq T_\ga$. The $\phi$
particles production starts then for temperatures $T$ below
$T_{gen} \propto g^2 $, where $\Gamma/H>1$. The energy scale at
which the $\phi$ is generated is fixed by the coupling constant
$g$, so the coincidence problem is explained in terms of the
strength of the interactions of the $\phi$ field. In particular
one can generate the $\phi$ field at low energies, e.g.
$E_{gen}\simeq 1\, eV$ with $g\simeq 10^{-12}$ giving a fine
structure constant $\alpha = g^2/4\pi\simeq 10^{-25}$ and a cross
section $\sigma\simeq g^2/32 \pi E_{gen}^2 \simeq 1 \, pb$. The
value of $\sigma$ is quite close to cross section of WIMP dark
matter with nucleons $\sigma_{w} \simeq  pb$ \ci{wimp}. Since we
can choose the coupling $g$ in such a way that the $\phi$ is
generated at low energies close to present time, this offers a new
interpretation of the cosmological coincidence problem:   dark
energy domination starts at such small energies because of  the
size of the coupling constant $g$. For a $\phi$-generation energy
of $E_{gen}\simeq 1 \, eV$ the ratio in energy densities from the
appearance of the fundamental field $\phi$ to present time is
$\rho_{DE}/\rho_{gen} = 10^{-12}$ and should be compared to a case
where $\phi$ is present at the Planck time $\rho_{pl}$ with
$\rho_{DE}/\rho_{pl} = 10^{-124}$, giving  a difference in ratios
of $112$ orders of magnitude. Therefore, the amount of fine tuning
in the parameters of the dark energy potential $v(\phi)$ is much
less sever in our case than in a standard quintessence dark energy
model. The production of the $\phi$ particles gives rise to the
scalar potential $v(\phi)$ and the field $\phi$ will then evolve
classically given by its equation of motion.  Of course, we still
need to choose $v(\phi)$ appropriately to give dark energy, for
example $v(\phi) = v_{DEo}/\phi$ with $v_{DEo}$ the present time
dark energy potential \cite{axQ}. We stress the fact that the dark
energy behavior of the universe is due to the form of the
potential $v(\phi)$ but the energy $E_{gen}$ at which the $\phi$
is generated is fixed by the coupling constant $g$.
\begin{figure}[tbp]
\begin{center}
\includegraphics*[width=7cm]{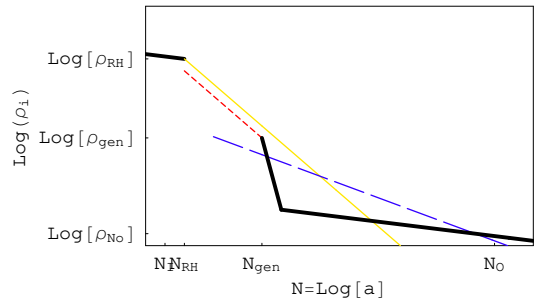}
\caption{We plot the logarithms of the energy densities $\rho_i$,
with the density of $\phi$ field (black line), $\vp$ field (red
dottedline),  radiation (yellow line) and of matter (blue dashed
line) against the number of e-folds $N=Log[a]$. We see how at
$N_{RH}$ the $\phi$ field disappears and the universe is reheated.
At $N_{gen}$ the $\phi$ field is re-generated and the $\varphi$
field disappears. We see how the $\phi$ field is rapidly diluted
and than maintains nearly constant at $\rho_\phi \simeq
\rho_{DE}$. At late times close to present time, i.e. at $N \simeq
N_o$, the $\phi$ field starts to dominate and inflate the
universe.} \label{figaaa}
\end{center}
\end{figure}
An interesting possibility is to unify inflation and dark energy
using the same scalar field $\phi$ in our dark energy quantum
generation picture \cite{axelfab}. We assume that $\phi$ has a
potential $v(\phi)$ which gives an early inflation at $E_I$, as in
standard inflationary models \ci{inflationarycosmology}. In this
case we use the coupling $L_{int} = g \, \phi \, \varphi^3$ to
allow the $\phi$ field to decay into $\varphi$ particles after
inflation, via the process $\phi \rightarrow \varphi + \varphi +
\varphi$. This decay process is very efficient \ci{kls2} and the
$\phi$ decays completely disappearing entirely from the spectrum
of the universe. In order to produce SM particles, the $\varphi$
is   coupled with the SM  via the standard interactions $L_{int} =
h \, \varphi^2 \, \chi^2$  or $L_{int} = \sqrt{h} \, \varphi \,
\bar\psi \psi$ with $\chi, \psi$ SM scalar or fermions,
respectively. Reheating of the early universe takes place via a
$2\leftrightarrow2$ process $\varphi + \varphi \leftrightarrow
\chi+\chi$ or $\varphi + \varphi \leftrightarrow \bar\psi+\psi$
with a transition rate $\Gamma_{RH}$, for $\Gamma_{RH}/H\equiv
T_{RH}/T_\vp\propto h^2/T_\vp>1$ at an energy scale $E_{RH}
\propto h^2$. Because of its couplings, the SM particles and
$\varphi$ are in thermal equilibrium at the time of reheating, and
as long as $\vp$ remains relativistic   one has $\Omega_\varphi
\simeq \Omega_{SM}\,g_\vp/g_{SM}$. After reheating $\rho_\vp$
evolves as radiation and eventually it will re-generate the $\phi$
field, at an energy $E_{gen} \propto g^2$. The re-generation
process of $\phi$ can be produced with the same interaction
$L_{int} = g \, \phi \, \varphi^3$ as the inflaton decay  at
$E_I$.  A main difference in the transition processes at $E_I$ and
at $E_{gen}$ is   the size of the mass of $\phi$. It varies quite
significantly and we should have $m_\phi(E_I)\gg m_\phi(E_{gen})$
and also $E_{gen}\gg m_\phi(E_{gen})$. In this case the
inflaton-dark energy potential $v(\phi)$ must be chosen to be flat
at high energy $E_I$, to give inflation, and at low energy $E_o$
to accelerate the universe at present time. The reheating energy
$E_{RH} \propto h^2$ and the $\phi$ generation energy $E_{gen}
\propto g^2$ are fixed by the coupling constants $h$ and $g$
independently of the potential $v(\phi)$. In the model presented
in section \ref{unification} we set $g=h^2 \propto E_I/m_{pl}$
reducing the number of parameters and connecting the inflation
scale to the $\phi$-generation and to the reheating scales,
$E_{gen} \propto E_I^2/m_{pl},\,  E_{RH} \propto E_I$. As
mentioned in the introduction, any unification of inflation and
dark energy, such as quintessential models,   need to explain how
the universe is reheated with SM particles and must account for
the long period of decelerating universe.  In our model we are
able to explain both features in a consistent way.

We give a schematic picture of the inflation-dark energy unified
model showing in fig.(\ref{figaaa}) the evolution of the
logarithms of the energy densities of the $\phi$ and  $\vp$
fields, radiation and matter. We start with a $\phi$ dominated
universe and then at $N_{RH}$, where $N\equiv \ln(a)$ with $a(t)$
the scale factor, the $\phi$ field decays and disappears,
reheating the universe. After reheating the universe is radiation
dominated, as in the standard cosmological model, and radiation
includes the SM relativistic degrees of freedom plus the extra
relativistic field $\varphi$. At $N_{gen}$ the $\phi$ field is
re-generated with $\rho_{\phi_f} \simeq \rho_{\varphi_i}$, where
$\rho_{\phi_f}$ is the value of $\rho_\phi$ at the end of the
generation process and $\rho_{\varphi_i}$ is the value of
$\rho_\varphi$ at the begin of the $\phi$ generation. At the same
time the $\varphi$ field decays and $\rho_{rel}$ diminishes. After
the generation, $\rho_\phi$ is rapidly diluted and subsequently
maintains nearly constant at $\rho_\phi \simeq \rho_{DE}$ and it
starts to dominate close to present times, i.e. at $N \simeq N_o$,
inflating the universe.

\section{The $\phi$ generation.}\label{regeneration}

In this section we describe  the quantum generation of the
quintessence scalar field $\phi$ and the subsequent appearance of
the dark energy behavior of the late universe due to the classical
evolution of $\phi$. In what follows we assume a universe that
consists of the particles of the standard model ''SM'' together
with  a massless  scalar field $\varphi$ and a quintessence scalar
field $\phi$ coupled together via a four particle interaction, as
for example $L_{int} = g \, \phi \, \varphi^3$ or $L_{int} = g \,
\phi^2 \, \varphi^2$. The $\varphi$ field is also coupled with the
SM through the interactions $\sqrt{h} \, \varphi \, \bar\psi \psi$
or $h \, \varphi^2 \, \chi^2$, where $\psi$ and $\chi$ are some SM
fermions or scalars respectively. We assume that at temperatures
above $1\, TeV$ all the SM particles and $\varphi$ are
relativistic and $\varphi$ is in thermal equilibrium with
$T_\varphi = T_{rad}$. As long as $\varphi$ is relativistic
$T_\varphi \simeq T_{rad}$ and $\Omega_\varphi \simeq
\Omega^{SM}_{rad} \, \frac{g_\vp}{g^{SM}_r}$, where
$\Omega^{SM}_{rad}$ is the SM radiation density and
$g_\vp,g^{SM}_r$ are the relativistic degree of freedom of $\vp$
and SM, respectively. We also suppose that at temperatures $T \gg
1 eV$  the universe does not contain any $\phi$ particles,
therefore the number density of $\phi$ particles is zero $n_\phi
=\Omega_\phi= 0$.

We will show that the field $\phi$ is generated via the
quantum transition $\varphi + \varphi \rightarrow \varphi + \phi$
or $\varphi + \varphi \rightarrow \phi + \phi$ with a transition
rate $\Gamma_{\varphi + \varphi \rightarrow \varphi + \phi
}=\Gamma_{\varphi + \varphi \rightarrow \phi + \phi }=
\langle\sigma_{gen}v\rangle \, n_\varphi \equiv \Gamma_{gen}$,
where $\sigma_{gen}= g^2/32 \pi E^2$ is the cross section for a $2
\leftrightarrow 2$ relativistic particle process, $v$ is the
relative velocity and $n_\varphi$ is the density number of
the $\varphi$ particles \ci{G}. This take place at energies below
$E_{gen} \simeq c_{gen} \, g^2 \, m_{pl}$ with $c_{gen}$ a
constant,  when $\Gamma_{gen}/H>1$. At the end of the $\phi$
generation one has $\Omega_{\phi_f} \simeq \Omega_{\varphi_i}$,
where $\Omega_{\varphi_i}$ is the $\varphi$ density at the begin
of the $\phi$ particles production, and also $\Omega_{\varphi_f}
\ll \Omega_{\varphi_i}$.
The production of relativistic particles $\phi$  becomes  a source
term for the generation of the scalar potential $v(\phi)$. Once
$v(\phi)$ has been produced the classical equation of motion gives
the evolution of $\phi$.  The acceleration of the universe is then
due to the form of scalar potential $v(\phi)$.

We consider  a system composed by two scalar fields $\phi$ and
$\varphi$ plus SM, in  a flat FRW metric,  with a density
lagrangian
\begin{equation} \label{lagrangian1}
\textit{L} = \frac{\partial_\mu\phi \partial^\mu \phi}{2}  +
\frac{\partial_\mu \varphi \partial^\mu \varphi}{2}  -
V_T(\phi,\varphi) + \textit{L}_{SM}
\end{equation}
were $\textit{L}$ is the SM lagrangian and $V_T(\phi,\varphi) =
v(\phi) + B(\vp) + v_{int}(\phi,\varphi)$,  $v(\phi)$ and
$B(\varphi)$ are the classical potentials of the two scalar fields
$\phi$ and $\varphi$ and $v_{int}(\phi,\varphi) = -
\textit{L}_{int}(\phi,\varphi)$, where $\textit{L}_{int}$ is the
interaction lagrangian. $\textit{L}_{int}$ plays a double role: it
affects the classical evolution of the two scalar fields and it
also originates  the quantum transitions between $\phi$ and
$\varphi$ particles that we use to generate the $\phi$ field at
late times. We divide the $\phi(t,x)$ field into a classical
background configuration $\phi_c(t)$ plus a perturbation
$\delta\phi(t,x)$ as $\phi(t,x) = \phi_c(t) + \delta\phi(t,x)$,
where $\delta\phi(t,x)$ corresponds to the quantum configuration
of the $\phi$ field ($\phi$ particles). The background $\phi_c(t)$
and the perturbation $\delta\phi(t,x)$ are usually taken as
independent variables. However we choose to take as the two
independent variables $\phi(t,x)$ and $\delta\phi(t,x)$. Moreover
we express $\delta\phi(t,x)$  in terms of the number density
$n_\phi$ of $\phi$ particles via the relation $n_\phi = E_\phi \,
\delta\phi^2$, see appendix \ref{appendix2}. We stress the fact
that when $n_\phi = 0$ one also has $\delta\phi(t,x) = 0$ that
implies $\phi(t,x) = \phi_c(t)$, so the scalar field $\phi$ is in
its classical configuration. Following the same argument we write
$\varphi(t,x) = \varphi_c(t) + \delta\varphi(t,x)$ and describe
the $\varphi$ field through the variables $\varphi(t,x)$ and
$n_\varphi  = E_\varphi \, \delta\varphi^2$. Let us concentrate on
the process of $\phi$ particles production in the case that both
$\phi$ and $\varphi$ particles are relativistic and the number
density of $\phi$ particles $n_\phi$ is initially zero. We use a
four particles interaction, such  as $\textit{L}_{int} = g \, \phi
\, \varphi^3$ or $\textit{L}_{int} = g \, \phi^2 \, \varphi^2$ and
we consider a $2\leftrightarrow 2 $ the processes, e.g. $\varphi + \varphi \leftrightarrow
\varphi + \phi$ or $\varphi + \varphi \leftrightarrow \phi +
\phi$. As mentioned previously, the $\varphi$ field is initially
thermalized and  it has a phase space distribution given by the
Bose-Einstein distribution $f_\varphi(E) = 1/(e^{E/T_\varphi}-1)$,
so we can  take all the $\varphi$ particles with the same energy
$E_\varphi = \bar r T_\varphi \simeq T_{rad}$. Since the $2
\leftrightarrow 2$ interaction process  conserves energy the
$\phi$ particles become in thermal equilibrium with $\vp$ and they
has the same energy $E_\phi = E_\varphi = E$. Moreover the two
scalar fields $\phi$ and $\varphi$ are relativistic so their
energy scales as $E = E_i /a(t)$.
Before the $\phi$ generation  the density number of $\varphi$
particles evolves as $n_\varphi = \zeta(3)T_\varphi^3/\pi^2
$, so the transition rate for the considered
$2\leftrightarrow 2$ processes of relativistic particles is
expressed in terms of the energy of the quantum particles as
$\Gamma_{gen} = \langle\sigma_{gen}v\rangle \, n_\varphi = (\zeta(3)/32\pi^3 \bar{r}^3) \, g^2 \, E $
 and therefore it scales as $\Gamma_{gen}  \sim
1/a(t)$.
As discussed in section \ref{motivations}, in order to have a late
time dark energy generation, it is fundamental to have a ratio
$\Gamma_{gen}/H$ that grows up with the expansion of the universe.
If we are in radiation domination  $H \sim a(t)^{-2}$ in in matter
domination $H \sim a(t)^{-3/2}$. This means that the
$\Gamma_{gen}$ should decrease more slowly than $1/a(t)^2$ during
radiation domination and than $1/a(t)^{3/2}$ during matter
domination, as in our case. Taking a flat FRW metric, the
evolution of our system is described by the following equations
(see eqs.(\ref{systemregeneration}) in appendix \ref{appendix2}
for details)
\bea \label{equation1}
\dot n_{\phi} + 3 H n_{\phi}
=  \, \tilde{\Gamma} \left( n_\varphi - n_\phi \right) + 2 \,
\sqrt{E \, n_{\phi}} \, \dot\phi& &
\\ \label{equation2}
\dot n_\varphi + 3 H n_\varphi=   -  \, \tilde{\Gamma} \left(
n_\varphi - n_\phi \right) + 2 \, \sqrt{E \, n_{\varphi}} \,
\dot\varphi& &
\\ \label{equation3}
\ddot{\phi} + 3 H \dot\phi + E^{3/2}  \sqrt{n_{\phi}}
+\partial_\phi v(\phi)+ \partial_\phi v_{int} (\phi,\varphi) = 0
\\ \label{equation4}
\ddot{\varphi} + 3 H \dot\varphi + E^{3/2}  \sqrt{n_{\varphi}}   +
\partial_\varphi B(\varphi) + \partial_\varphi v_{int}(\phi,\varphi) = 0& &
\eea
together with the  Friedman equation $ H^2 =
\frac{1}{3}\rho_{T}$, where $\rho_T = \rho_\phi + \rho_\varphi +
\rho_{rad} + \rho_{mat}$, and  $\rho_{rad}$ and $\rho_{mat}$ are
the energy densities of radiation and matter respectively.
Moreover $v(\phi)$ and $B(\varphi)$ are the classical potentials
of the $\phi$ and $\varphi$ fields respectively and
$v_{int}(\phi,\varphi)$ is the classical interaction
potential.

Eqs.(\ref{equation1}) and (\ref{equation2}) are the Boltzmann
equations that govern the dynamic of the number densities $n_\phi$
and $n_\varphi$. They take into account the quantum transition
between $\phi$ and $\varphi$ particles thanks to the terms
proportional to $\tilde{\Gamma}$. The quantity $\tilde{\Gamma}$ in eqs. (\ref{equation1}) and
(\ref{equation2}) is given by  $\tilde{\Gamma} \equiv
\langle\sigma_{gen}v\rangle n_\varphi$ for the process $\varphi
\varphi \leftrightarrow \varphi \phi$ and $\tilde{\Gamma} \equiv
\langle\sigma_{gen}v\rangle (n_\varphi+n_\phi)$ for the process
$\varphi \varphi \leftrightarrow \phi \phi$. $\tilde{\Gamma}$
takes into account the
contribution of the two processes $\varphi \varphi
\rightarrow \varphi \phi$ and its inverse  $\varphi \phi \rightarrow \varphi
\varphi$ in one case and
$\varphi \varphi \rightarrow \phi \phi$
and $\phi \phi \rightarrow \varphi \varphi$ in the other case, since
one has $
\Gamma_{\varphi+\varphi \rightarrow \varphi + \phi} \, n_\varphi
-\Gamma_{\varphi+\phi \rightarrow \varphi + \varphi} \, n_\phi =
\langle\sigma_{gen}v\rangle \, n_\varphi \,(n_\varphi - n_\phi)=\tilde{\Gamma} \left( n_\varphi - n_\phi \right)$
for  $\varphi+\varphi \leftrightarrow \varphi + \phi$
and $
\Gamma_{\varphi+\varphi \rightarrow \phi + \phi} \, n_\varphi
-\Gamma_{\phi+\phi \rightarrow \varphi + \varphi} \, n_\phi =
\langle\sigma_{gen}v\rangle \,(n_\varphi^2 - n_\phi^2)=\tilde{\Gamma} \left( n_\varphi - n_\phi \right) $ for
$\varphi+\varphi \leftrightarrow \phi + \phi$.
At the
begin of the $\phi$ generation one has $n_\phi = 0$ and therefore
$\tilde \Gamma = \Gamma_{gen}$ in both processes.
Moreover the two terms $2 \sqrt{E \, n_\phi} \, \dot\phi$ and $2
\sqrt{E \, n_\varphi} \, \dot\varphi$ contained in
eqs.(\ref{equation1}) and (\ref{equation2}) respectively, couple
the number densities $n_\phi$ and $n_\varphi$ with the scalars
$\phi$ and $\varphi$ respectively. Eqs.(\ref{equation3}) and
(\ref{equation4}) are the equations of motion of the lagrangian in
eq.(\ref{lagrangian1}) for the scalar fields $\phi(t,x)$ and
$\varphi(t,x)$. These equations contains the terms $E^{3/2}
\sqrt{n_\phi}$ and $E^{3/2} \sqrt{n_\varphi}$ that couple the two
scalar fields $\phi$ and $\varphi$ with their number densities
$n_\phi$ and $n_\varphi$ and become the source terms for
generating the scalar potential $v(\phi)$. Note that the terms
$E^{3/2} \sqrt{n_{\phi}}$ and $E^{3/2} \sqrt{n_{\phi}}$ contained
in eqs.(\ref{equation3}) and (\ref{equation4}) represent the
spatial derivatives of the two scalar fields, in fact they come
out from the relations $-\frac{\nabla^2 \phi}{a(t)} = E^{3/2}
\sqrt{n_\phi}$ and $-\frac{\nabla^2 \varphi}{a(t)} = E^{3/2}
\sqrt{n_\varphi}$ , see eq.(\ref{nablaphi}) in appendix
\ref{appendix2}. Of course eqs.(\ref{equation1})-(\ref{equation4})
conserve energy-momentum. The energy density and pressure of the
$\phi$ field are $\rho_\phi = \rho_{1 \phi} + \rho_{2 \phi}$ and
$p_\phi = p_{1 \phi} + p_{2 \phi}$, where $\rho_{1 \phi} \equiv
\fr{\dot\phi^2}{2} + v(\phi)$, $\rho_{2
\phi} \equiv E \, \frac{n_\phi}{2}$, $p_{1 \phi} \equiv
\frac{\dot\phi^2}{2} - v(\phi)$ and $p_{2
\phi} = \rho_{2 \phi}/3$ (see appendix \ref{appendix2}).
In the same way the energy density and the pressure of the
$\varphi$ field are $\rho_\varphi = \rho_{1 \varphi} + \rho_{2
\varphi}$ and $p_\varphi = p_{1 \varphi} + p_{2 \varphi}$, where
$\rho_{1 \varphi} \equiv \frac{\dot\varphi^2}{2} + v_{int}(\phi,\varphi)+B(\varphi)$,
$\rho_{2 \varphi} \equiv E \, \frac{n_\varphi}{2}$, $p_{1 \varphi}
\equiv \frac{\dot\varphi^2}{2} - v_{int}(\phi,\varphi)-B(\varphi) $ and $p_{2 \varphi} =
\rho_{2 \varphi}/3$. It is easy to show (see appendix
\ref{appendix2}) that
\bea\label{energycondition4}
\dot\rho_\phi + 3 H (\rho_\phi + p_\phi)  = \frac{1}{2} E \,
\tilde{\Gamma} \left( n_\varphi - n_\phi \right) - \dot\phi \,
\partial_\phi v_{int}
\\
\dot\rho_\varphi + 3 H (\rho_\varphi + p_\varphi)  = - \frac{1}{2}
E \, \tilde{\Gamma}  \left( n_\varphi - n_\phi \right)
+\dot\phi \, \partial_\phi v_{int}
\eea
Therefore eqs.(\ref{energycondition4}) give the
energy-momentum conservation law $\dot\rho_T + 3 H (\rho_T + p_T)
= 0$, where $\rho_T = \rho_\phi + \rho_\varphi$ and $p_T = p_\phi
+ p_\varphi$.
Now we are ready to describe the $\phi$ particles production
qualitatively. For presentation purposes  we will not take into
account the expansion of the universe, i.e. we will take $H = 0$,
however we show below that the growth of the scale factor is very
small (of the order of $H/\Gamma_{gen}$)
\footnote{A way to account
for the evolution of the universe is to make the following
substitutions $\dot\phi = E^3 \dot{\tilde\phi}$, \, $n_\phi = E^3
\tilde{n}_\phi$ and $n_\varphi = E^3 \tilde{n}_\varphi$, where $E
= E_\varphi = E_\phi = E_i/a(t)$ in such a way that the EoM read
$\dot{\tilde{n}}_{\phi} =  - \tilde{\Gamma} \left( \tilde{n}_\phi
- \tilde{n}_\varphi \right) +E_{\phi}^{2} \sqrt{\tilde{n}_{\phi}}
\, \dot{\tilde\phi}$, \, $\dot{\tilde{n}}_\varphi  =
\tilde{\Gamma} \left( \tilde{n}_\phi  - \tilde{n}_\varphi \right)$
\, $\ddot{\tilde\phi} + \sqrt{\tilde{n}_{\phi}} +
\partial_\phi v(\phi)/E^3_\phi = 0$.
These equations are totally equivalent to the system
(\ref{equationbis1}-\ref{equationbis3}) if $a(t) \simeq a_i$
during the $\phi$ regeneration.}.
Moreover in this presentation we will not consider
the contribution from classical interactions and we also take $B(\varphi)=0$. We will
also work with  $\tilde{\Gamma}$ constant. Obviously
$\tilde{\Gamma}$ depends both on the energy of the decaying
particles and on the number densities, but we use this
approximation to write an  analytical solution of
eqs.(\ref{equation1})-(\ref{equation4}). Of course comparison of
the approximated analytical solution with the numerical solution
shows a complete agreement and the classical interaction
$v_{int}(\phi,\varphi)$ and expansion rate do not play
a significant role. Under these approximations the system of
eqs.(\ref{equation1}-\ref{equation4}) reduce to  the following
equations
\bea
\label{equationbis1}
\dot n_{\phi}  &=& \tilde{\Gamma} \left(
n_\varphi  - n_\phi \right) + 2 \, \sqrt{ E_i \, n_{\phi}} \,
\dot\phi\\
\label{equationbis2} \dot n_{\varphi}  &=& -
\tilde{\Gamma} \left( n_\varphi  - n_\phi
\right) + 2 \, \sqrt{ E_i \, n_{\varphi}} \, \dot\varphi\\
\ddot{\phi} &+&  E_i^{3/2} \sqrt{n_{\phi}} +
\partial_\phi v(\phi) = 0\\
\label{equationbis3}\ddot{\varphi} &+&  E_i^{3/2}
\sqrt{n_{\varphi}}  = 0
\eea
where $E_i$ is the energy of the quantum particles at the begin of
the $\phi$ generation. The initial conditions on the $\varphi$
field are $\rho_{\varphi_i}= E_i n_{\varphi_i}$, $n_{\varphi_i} = \frac{\varsigma(3)}{\pi^2 \bar r^3}
E_i^3$ with  $ \langle \dot\varphi_i^2\rangle = E_i \, n_{\varphi_i}$
and   $\Omega_{\varphi_i} = 0.1$. For the $\phi$ field we have the initial
conditions $\rho_{\phi_i}=0$ that gives  $\Omega_{\phi_i} =
n_{\phi_i} = \dot\phi_i =v(\phi_i) = 0$.

Now, the first stage starts with
$\left(n_{\varphi} - n_\phi\right) \tilde{\Gamma} \gg - 2
\sqrt{E_i \, n_{\phi}} \, \dot\phi$ in eq.(\ref{equationbis1}),
which is clearly verified at the begin since $n_{\phi_i}=0$, and the
approximate solution of  eqs.(\ref{equationbis1}-\ref{equationbis3}) is
\bea\label{sol1}
n_\phi &\simeq& n_{\varphi_i}\, \tilde{\Gamma}
\, \left(t-t_i\right) \left[1-\frac{4}{9} E_i^2 \left(t-t_i\right)^2 \right],\nonumber\\
n_\varphi &\simeq& n_{\varphi_i}
\left[1+ (2E_i -\tilde{\Gamma}) \left(t-t_i\right) \right], \\
\dot\phi &\simeq& - \frac{2}{3} \, E_i^{3/2} \,
\sqrt{n_{\varphi_i}\tilde{\Gamma}} \; \, \left(t-t_i\right)^{3/2}, \nonumber\\
\dot\varphi &\simeq& \dot\varphi_i \left[1-E_i \left(t-t_i \right)
\right]. \nonumber \eea
valid for $\tilde{\Gamma} \,
\left(t-t_i\right)\ll 1$ and $E_i \, \left(t-t_i\right)\ll 1$. This
first phase of the generation consists in the growth of $n_\phi$
due to $\phi$ particle production. However the growth of $n_\phi$
ends at some time $t_1$, when $\left(n_{\varphi} - n_\phi\right)
\tilde{\Gamma} = - 2 \sqrt{E_i \, n_{\phi}} \, \dot\phi>0$ and $\dot
n_\phi = 0$. This point is the end of stage one and  $n_\phi$
reaches a maximum  value
$n_\phi = \bar{n}_\phi \equiv n_\varphi + \beta ( 1 -(1
+ 2 n_\varphi/\beta)^{1/2})$ with $\beta = 2 E_i
\dot\phi^2/\tilde{\Gamma}^2$.
At the end of this first stage one also has
$n_\varphi/\beta \ll 1$, $n_{\varphi}(t_1) \simeq n_{\varphi_i}$, $\dot\varphi(t_1) \simeq
\dot\varphi_i$ and $n_\phi(t_1)=\bar{n}_\phi
\simeq n_\varphi^2/2\beta \ll n_{\varphi}$.

The second phase of the generation maintains $\left(n_{\varphi} -
n_\phi\right) \tilde{\Gamma}  + 2 \sqrt{E_i \, n_{\phi}} \,
\dot\phi = 0$ dynamically, therefore from eq.(\ref{equationbis1})
it follows that $n_\phi$ remains constant at the equilibrium value
$\bar{n}_\phi \ll n_{\varphi_i}$. One can cheek the stability of
this solution $n_\phi = \bar{n}_\phi$ analyzing the dynamical
equation for small perturbations of $n_\phi$ around $\bar{n}_\phi$
that is $\delta\dot n_\phi = - \beta \, \delta n_\phi$ with $\beta
= \tilde{\Gamma} -
\sqrt{\frac{E_i}{\bar{n}_\phi}}\frac{\dot\phi}{2}
> 0$, that gives an exponential suppression of perturbations, so
the solution $n_\phi = \bar{n}_\phi$ is stable. If $n_\phi$ is
maintained constant at its equilibrium value $\bar{n}_\phi$, the
relation $2 \sqrt{E_i \, n_\phi} \, \dot\phi = - \tilde{\Gamma} \,
\left( n_\varphi - \bar{n}_\phi \right) =\dot n_\varphi$ is
satisfied dynamically (see eqs.(\ref{equationbis1}) and
(\ref{equationbis2})), so we can write eq.(\ref{equationbis3}) in
the form $\ddot{\phi} +
\partial_\phi v(\phi) \simeq \frac{1}{2} \, E_i \, \tilde{\Gamma} \, \frac{n_\varphi -
\bar{n}_\phi}{\dot\phi}$, or $\dot{\rho}_{1\phi} = \frac{1}{2} \,
E_i \,\tilde{\Gamma} \, \left(n_\varphi - \bar{n}_\phi \right)$
where $\dot{\rho}_{1\phi} \equiv \frac{\dot\phi^2}{2} + v(\phi)$.
 We take  $\dot\varphi^2 \simeq E\,n_\varphi$, which agrees well
 with the numerical simulation, and
$E\, \dot\varphi \, \sqrt{E n_\varphi} \simeq \, E \, \dot
\varphi^2 \simeq 2 \, E \, \rho_{1\varphi}$ where  $
\rho_{1\varphi} \equiv \dot\varphi^2/2$.
Therefore
eqs.(\ref{equationbis1})-(\ref{equationbis3}) are reduced to the
following effective equations
\bea
\label{effectiveequatins} {n}_\phi = \bar{n}_\phi, \qquad
\dot{n}_\varphi &=&  -\tilde{\Gamma}
\left(n_\varphi - \bar{n}_\phi \right)+ 4\, \rho_{1\varphi}, \nonumber \\
\dot{\rho}_{1\phi} &=& \frac{1}{2} \tilde{\Gamma}\, E_i \,
\left(n_\varphi - \bar{n}_\phi \right),\\
\dot{\rho}_{1\varphi} &=& -2 \, E \,\rho_{1\varphi}\nonumber
\eea
whose solutions are
\begin{equation}
\begin{array}{ll}
\label{sol2} n_\phi = \bar{n}_\phi,\\
n_\varphi =n_{\varphi_i} \, \left[ e^{-\tilde{\Gamma} (t-t_1)} +
2E_i \, \frac{  e^{-\tilde{\Gamma} (t-t_1)}-e^{-2E_i(t-t_1)}}{2E_i-\tilde{\Gamma}} \right]+ \\
\qquad \,\, +\,\, \bar{n}_\phi \left[1-e^{-\tilde{\Gamma}(t-t_1)}\right],\\
\rho_{1\phi} = \frac{E_i}{2} (n_{\varphi_i}-\bar{n}_\phi)
\left[1-e^{-\tilde{\Gamma}(t-t_1)}\right] +\\
\qquad \,\,+\,\, E_i n_{\varphi_i} \, \frac{2E_i
\left(1-e^{-\tilde{\Gamma}(t-t_1)}\right) -\tilde{\Gamma}\left(1-e^{-2E_i(t-t_1)}\right)}{2(2E_i-\tilde{\Gamma})},\\
\rho_{1\varphi} = \frac{E_i n_{\varphi_i}}{2} e^{-2E_i(t-t_1)}.
\end{array}
\end{equation}
From these equations we see that $n_\varphi$ decrease
 until it reaches the equilibrium at $n_{\varphi_f} \simeq n_{\phi_f}
\simeq \bar{n}_\phi \ll n_{\varphi_i}$ and that $\rho_{1\varphi}$
decrease exponentially to zero so at the end of the regeneration
one has $\rho_{\varphi_f} = E_i \bar{n}_\phi \ll
\rho_{\varphi_i}$. Moreover $\rho_{1\phi}$ grows until
$\rho_{1\phi_f} =\rho_{\varphi_i}- E_i \bar{n}_\phi$ so one has
$\rho_{\phi_f} =\rho_{\varphi_i}$ and in conclusion one has
$\Omega_{\phi_f} \simeq \Omega_{\varphi_i}$, $\Omega_{\varphi_f} \ll
\Omega_{\varphi_i}$ and $v(\phi) \simeq \rho_{\varphi_i}$, that
means that all the energy initially stored into the quantum field
$\varphi$ has been transferred to the $\phi$ field in accordance
with the total energy conservation.

\begin{figure}[tbp]
\begin{center}
\includegraphics*[width=7cm]{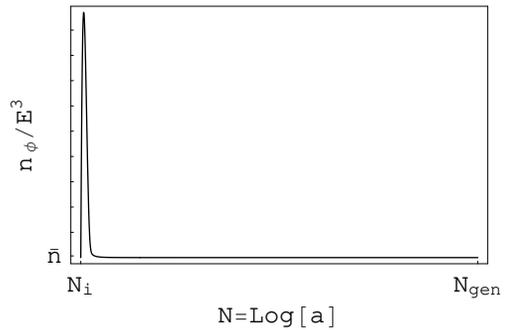} \caption{ We show the
evolution  of  $n_\phi/E^3$ against the number of e-folds $N$
during the $\phi$ generation in the case $v(\phi) = v_i/\phi$. We
see how $n_\phi$ grows from zero and than reaches the equilibrium
value $\bar{n}_\phi$.} \label{fign1}
\end{center}
\end{figure}

\begin{figure}[tbp]
\begin{center}
\includegraphics*[width=7cm]{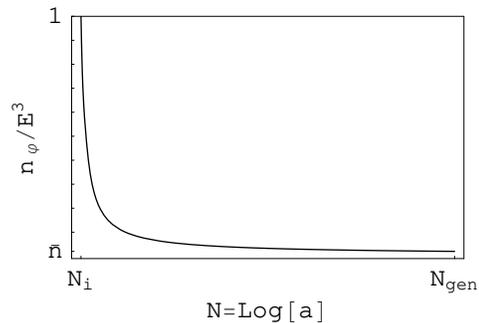} \caption{ We show the
evolution  of  $n_\varphi/E^3$ against the number of e-folds $N$
during the $\phi$ generation in the case $v(\phi) = v_i/\phi$.
$n_\varphi$ decreases exponentially and reaches the asymptotic
value $\bar{n}_\phi$.} \label{fign2}
\end{center}
\end{figure}

We solve numerically the system of
eqs.(\ref{equation1}-\ref{equation4}) in the case $v(\phi) =
v_i/\phi$.  In fig.(\ref{fign1}) we show the evolution of
$n_\phi/E^3$ and in fig.(\ref{fign2}) that of $n_\varphi/E^3$. The
evolution of $n_\phi$ includes the solutions given in
eqs.(\ref{sol1}) and (\ref{sol2}) showing that $n_\phi$ starts
from zero, it reaches its maximum and then it decreases to its
equilibrium value $\bar{n}_\phi$. The evolution of $n_\varphi$ is
plotted in fig.(\ref{fign2}) and we see how $n_\varphi$ decrease
exponentially to the asymptotic value $\bar{n}_\phi$ as in
eqs.(\ref{sol1})and (\ref{sol2}). In fig. (\ref{fig5}) we plot the
evolution of the density parameter $\Omega_{\varphi}$ showing that
$\Omega_{\varphi}$ goes to zero at the end of the $\phi$
generation.
\begin{figure}[tbp]
\begin{center}
\includegraphics*[width=7cm]{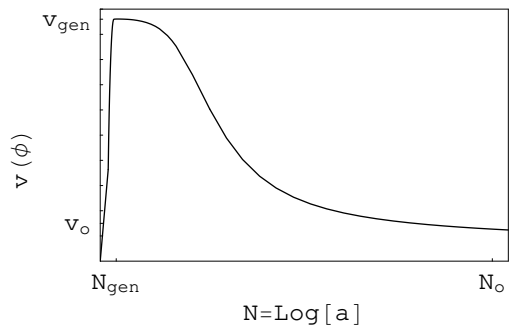} \caption{We show the
evolution  of  the potential $v(\phi)$  against the number of
e-folds $N$ in the case $v(\phi) = v_i/\phi$. } \label{figv}
\end{center}
\end{figure}
\begin{figure}[tbp]
\begin{center}
 \includegraphics*[width=7cm]{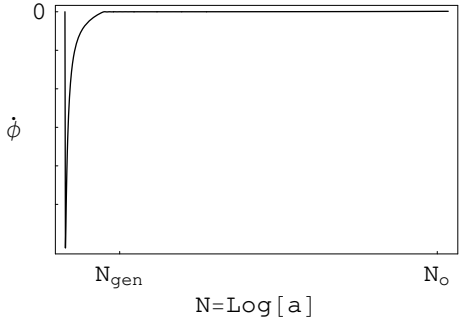}
\caption{We show the evolution  of  the
derivative $\dot\phi(N)$ against the number of e-folds $N$ in the
case $v(\phi) = v_i/\phi$.} \label{figdv}
\end{center}
\end{figure}
\begin{figure}[tbp]
\begin{center}
\includegraphics*[width=7cm]{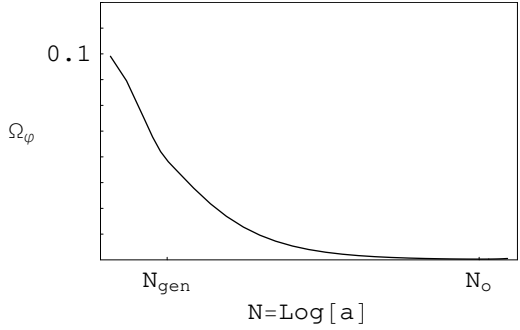}
\caption{We show the evolution  of  $\Omega_{\varphi}$ against the
number of e-folds $N$. We see that $\Omega_{\varphi}$ rapidly goes
to zero at the end of the $\phi$ genearion.} \label{fig5}
\end{center}
\end{figure}
We stress some important features of the $\phi$ generation
process. The first one is that, in spite of starting with a
$v(\phi_i) = 0$, at the end of the process one has generated a
potential $v(\phi) \simeq \rho_{\varphi_i} \neq 0$. In
fig.(\ref{figv}) we show how $v(\phi)$ grows from zero to its
maximum value $v(\phi) \simeq \rho_{\varphi_i}$. At this point the
quantum generation process is completed and the subsequent
evolution of $\phi$ is given by its classical equations of motion.
At the same time in fig.(\ref{figdv}) we see the evolution of
$\dot\phi$, where it starts at $\dot\phi(t_{gen i}) = 0$ and takes
negative values (implying the growth of $v(\phi)$) and eventually
reaches positive values. The time when $\dot\phi(t_{gen f}) = 0$
corresponds to the end of $\phi$ generation and to the maximum of
$v(\phi)$. Another feature of the $\phi$ generation process
consists of the existence of an equilibrium value $\bar{n}_\phi
\ll n_{\varphi_i}$ for the density number of $\phi$ particles
$n_\phi$. Therefore after a first phase of growth, $n_\phi$
saturates at $\bar{n}_\phi$ and then it is maintained constant.
This means that all the energy coming from further $\varphi$
particles decay  is transferred directly into the potential
$v(\phi)$ without generate any change in $n_\phi$.

Let us point out more properties of the $\phi$ generation process.
First we note that if the $\phi$ field is regenerated when the
universe is radiation dominated, the scale factor evolves as $a(t)
= a_i \sqrt{1 + H_i \left(t-t_i\right)}$.
The $\phi$ generation ends at $\tilde \Gamma (t-t_i) \gtrsim
1$ (see eq.(\ref{sol2})) with $\tilde \Gamma \simeq \Gamma_{gen}$,
giving $H (t-t_i) \lesssim H/\Gamma_{gen} \leq 1$ and $ a(t)
\lesssim \sqrt{2} a_i$ (the generation process has $\Gamma_{gen}/H
\lesssim 1$), thus we see that the expansion of the universe
during the $\phi$ generation plays no significant role.
We also stress the fact that the generation process preserves the
homogeneity and isotropy  of the universe, as in the reheating
process after inflation \ci{inflationarycosmology}. In fact it is obtained via a $2
\leftrightarrow 2$ quantum process in which the energy-momentum is
conserved ($\sum p_i = \sum p_f$) so the amount of homogeneity
is preserved.

Now we can state something more about the begin and the
duration of the generation process. First note that at the begin
of the $\phi$ generation $n_\phi=0$, that implies $\tilde\Gamma =
\Gamma_{gen}\equiv \Gamma_{\varphi+\varphi\rightarrow
\phi+\varphi}=\Gamma_{\varphi+\varphi\rightarrow \phi+\phi}$.
Therefore, in order to establish the efficiency of $\phi$
particles production one should compare the transition rate
$\Gamma_{gen}$ with the expansion rate of the universe $H$. We
assume that before the generation the $\varphi$ field has the same
temperature of the radiation, so we take $E_{rad} = E_\varphi =
E$.  If the generation  starts before the matter domination one
has $H = (\rho_{rel}/3 \, m_{pl}^2\Omega_{rel})^{1/2}  \equiv   c_H
E^2$ where $c_H \equiv (\pi^2 g_{rel}/90 \, \bar{r}^4
\, m_{pl}^2 \, \Omega_{rel})^{1/2}$ and $g_{rel}$ is the total number
of relativistic degrees of freedom. Therefore, using $n_\varphi =
\frac{\zeta(3)}{\pi^2}\,T_\varphi^3$ before the $\phi$ generation
one has $\Gamma_{gen} = \frac{\zeta(3)}{32 \pi^3 \bar{r}^3} \, g^2
\, E$ and $\Gamma_{gen}/H = c_{gen} \, g^2 \, m_{pl} / E$, where
we have defined $c_{gen} \equiv (\zeta(3)/32 \pi^4 \bar{r})
\, (90 \, \Omega_{rel}/g_{rel})^{1/2}$.
We conclude that the $\phi$ generation starts at energies  $E
\lesssim E_{gen}$ with
\begin{equation}\label{egen}
E_{gen} \equiv c_{gen} \, g^2 \, m_{pl}=1.6 \left(\fr{g}{10^{-12}}\ri)^2 \,eV
\end{equation}
when $\Gamma_{gen}/H \equiv E_{gen}/E \gtrsim 1$. We aim to have a
$\phi$ particles production at energy scales well below  $MeV$,
where the number of relativistic degrees of freedom is $g_{rel} =
g^{SM}_{rel} + g_\vp \simeq 4.36$ giving $c_{gen} \simeq 6 \cdot
10^{-4}$. Of course, by means of eq.(\ref{egen}) we can
conveniently choose the coupling $g$ in such a way that the
generation process starts at an energy of about $1 eV$. This takes
place if $g \simeq 10^{-12}$ or $\alpha = g^2/4\pi \simeq
10^{-25}$, where $\alpha$ is the fine structure constant of the
transition process. We stress the fact that
$\Gamma_{gen}/H \equiv E_{gen}/E \geq 1$ during  the $\phi$
generation process but, when the $\phi$ field starts to inflate
the universe, one  has $H \sim
a(t)^{-\frac{3}{2}(1+\omega_{\phi})}$ with $\omega_\phi \simeq
-1$, so $H$ will be roughly constant and at the dark energy epoch
$\Gamma_{gen}/H \sim a(t)^{\frac{3}{2}(1+\omega_\phi)-1}
\rightarrow 0$. Therefore,  $\phi$ and $\varphi$ fields  decouple
at $E_{dec} = \frac{\sqrt{3}}{8\pi} \frac{\sqrt{\rho_{DE}}}{g^2 \,
m_{ Pl}} = \frac{\sqrt{3}\, c_{gen}}{8\pi}
\frac{E_{DE}^2}{E_{gen}}$.

We can summarize the $\phi$ generation process as follows. At the
begin of the generation we have $n_{\varphi i} =
\frac{\zeta(3)}{\pi^2\bar r^3} \, E_{\varphi i}^3$, \, $n_\phi =
\Omega_\phi =v(\phi)= 0$ and  at $E_{\vp i} =  E_{gen}$ we have
$\Gamma_{gen}=H$. The first effect is the decay of $\vp$ and the
growth of $n_\phi$ to its equilibrium value $\bar{n}_\phi \ll
n_{\varphi_i}$ after which $n_\phi$ remains constant. Subsequently
all the energy coming from the $\varphi$ field is transferred to
the potential $v(\phi)$ through the chain reaction $n_\varphi
\rightarrow n_\phi \rightarrow v(\phi)$. In this process $n_\phi$
remains constant and $\bar{n}_\phi$ represent a maximum transfer
efficiency value for $n_\phi$ in such a way that the energy coming
from the decaying $\varphi$ particles is immediately stored into
the potential $v(\phi)$. At the end of the generation on has
$n_{\phi_f} \simeq n_{\varphi_f} \simeq \bar{n}_\phi \ll
n_{\varphi_i}$ $\rho_{\varphi_f} \ll \rho_{\varphi_i}$ and
$\rho_{\phi_f} \simeq \rho_{\varphi_i}$.

\begin{figure}[tbp]
\begin{center}
\includegraphics*[width=7cm]{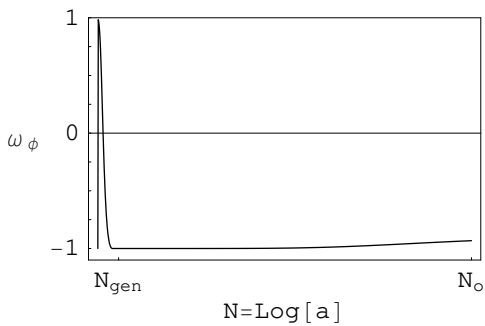} \caption{We show the
evolution  of the EoS parameter $\omega_\phi$ against the number
of e-folds N.} \label{figomega}
\end{center}
\end{figure}

\begin{figure}[tbp]
\begin{center}
\includegraphics*[width=7cm]{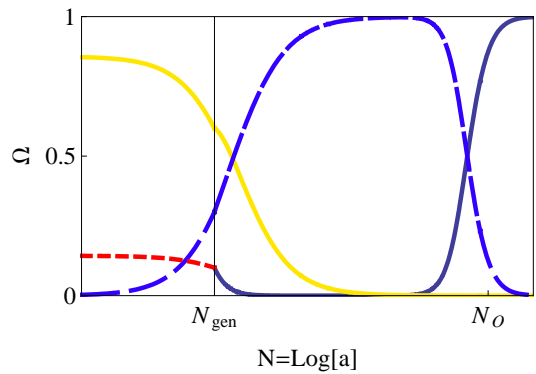} \caption{We plot  the
density parameters $\Omega_\phi$ (black line), $\Omega^{SM}_{rad}$
(yellow line), $\Omega_{mat}$ (blue dashed line) and
$\Omega_{\varphi}$ (red dotted line) against the number of e-folds
$N=Log[a]$. We see how at $N_{gen}$ the $\varphi$ disappears and
the $\phi$ field is generated. Present times are at $N_o$ when
$\Omega_\phi \simeq 0.74$ and $\Omega_m \simeq 0.26$.}
\label{fig4}
\end{center}
\end{figure}

From this point on, the evolution of $\phi$ is the standard one,
where its dynamics  is dominated by the classical potential
$v(\phi)$. The generation process described before does not depend
explicitly on  the form of the potential $v(\phi)$ and the value
of $\phi$ at the end of the generation process  is given by the
condition $v(\phi_{gen}) \simeq \rho_{\varphi_i} \simeq E_{\vp
i}^4=E_{gen}^4$. Since we want the $\phi$ field to give dark
energy, we must choose a potential $v(\phi)$ that slow rolls at
late times, i.e. at present time when $\phi \simeq \phi_o$.
Clearly we must impose the condition $v(\phi_{gen}) > v(\phi_o) =
v_{DE} \simeq (10^{-3} \, eV)^4$. For example one can consider an
effective potential $v(\phi) = \frac{v_i}{\phi}$ that verifies the
slow roll conditions for $\phi \geq  \sqrt{2}$ and take $v_i =
(10^{-3} eV)^4$. In this case the value of $\phi$ at the end of
the generation is $\phi_{gen} \simeq v_i/E_{gen}^4$ where we have
used $E_{gen}^4=v(\phi_{gen}) = \frac{v_i}{\phi_{gen}}$.
Typically, runaway quintessence potentials have an EoS parameter
$\omega_\phi$ that reaches positive values of $\omega_\phi \simeq
1$ diluting $\Omega_\phi$ and later there is a transition from
$\omega_\phi = 1$ to $\omega_\phi = -1$ as in fig.(\ref{figomega})
where $\Omega_\phi$ starts growing \ci{axQ}. Therefore the matter
domination epoch is unchanged by our $\phi$ generation scheme. In
fig.(\ref{fig4}) we show the evolution of the density parameters
$\Omega_\phi$, $\Omega^{SM}_{rad}$ and $\Omega_{mat}$.  The $\phi$
is generated  at $N_{gen}$ at energies $ E_{gen} \simeq 1 eV$, for
$g\simeq 10^{-12}$,  close to radiation matter equality with
$\Omega_{\phi_i} \simeq 0.2$ and present time is at $\Omega_\phi
\simeq 0.74$ and $\Omega_m \simeq 0.26$ with $\omega_\phi \simeq
-1$.

The cross section for the generation process is given by
\be\label{sg}
\sigma_{gen}=\fr{g^2}{32\pi E^2_{gen}} =
1.5 \left(\fr{10^{-12}}{g}\ri)^2 pb = 2.4 \left(\fr{1\,eV}{E_{gen}}\ri) pb
\ee
where we have used $E_{gen} \equiv c_{gen} g^2 \, m_{pl}$ and
$c_{gen} \equiv \frac{\zeta(3)}{32 \pi^4 \bar{r}} \,
\sqrt{\frac{90 \, \Omega_{rel}}{g_{rel}}}\simeq 6\cdot 10^{-4}$
and $pb=10^{-36}cm^2$.
It is interesting to compare the cross section $\sigma_g$ with that
of WIMPS. The relic abundance of WIMPS is
$\Omega_w h^2=3\times 10^{-27}cm^3/<\sigma_{w}v>$ \ci{wimp}
giving $\langle \sigma_{w}v\rangle=0.9\,c\, pb $, with $c$ the speed
of light. If we take that at decoupling
the WIMPS  have a mass to temperature ratio $m/T=20$ \ci{wimp} we
obtain $ \langle\sigma_{w}\rangle=2.4 pb$ equivalent to our $\sigma_{gen}$ for
$E_{gen}=1\,eV$. However,  the present time  observational
upper limit to $\sigma_w$  between WIMPS and nucleons is $\sigma_w \lesssim 10^{-42}cm^2$
consistent with supersymmetric WIMPS \ci{wimp}.

\section{Unification of inflation and dark energy.} \label{unification}

In this section we discuss the possibility of unifying inflation
and dark energy  by means of a unique scalar filed $\phi$ that we
call uniton, as in \cite{axelfab}. In general it is not difficult
to choose the  potential $v(\phi)$ in such a way that the $\phi$
field is responsible for both inflation and dark energy
\cite{scalarunification}. To achieve inflation and dark energy
with the same scalar $\phi$ one requires that the  potential
$v(\phi)$ must satisfy the slow roll conditions
$|v'(\phi)/v(\phi)| <1$ and $|v''(\phi)/v(\phi)| <1$ at high
energies for inflation and at low energies for dark energy.
Inflationary potentials which  have a minimum $v_{min} = 0$ at a
finite value of $\phi$  are not useful to unify inflation and dark
energy, since this kind of potentials do not inflate at low
energies \ci{axQ}. This kind of potentials may be useful to unify
inflation and dark matter. In most inflation-dark energy
unified models only the classical evolution of the quintessential scalar field is
considered and the reheating and the  long
period of decelerating universe (between inflation and dark energy)
are not taken into account.

In this section we present an inflation-dark energy unified scheme
that can be resumed in the following way: as in usual inflationary
models, the early universe is dominated by the $\phi$ field that
inflates for a sufficient number of e-folds. After the end of
inflation the $\phi$ field decays completely into the extra
relativistic field $\varphi$ already introduced in section
\ref{regeneration}.  The $\vp$ couples and produces SM particles
at an energy $E_{RH}$ and the universe is reheated. At low
energies $E_{gen}$ the $\phi$ field is generated via the quantum
generation mechanism studied in section \ref{regeneration} and the
universe enters the dark energy epoch at temperatures close to
present time. From $E_{RH}$ to $E_{gen}$ we have the standard
evolutionary scenario with the extra relativistic degree of
freedom $\varphi$.

The new feature in the unification scheme that we present here, is
that the transition between decelerating radiation-matter dominate
universe and the dark energy era is due to a quantum process, i.e.
the low energy generation of the $\phi$ field. As stated in
section \ref{regeneration}, the energy scale $E_{gen}$ at which the uniton $\phi$
is generated is fixed by the couplings $g$  of $\phi$.
The scales $E_{gen}$
may be many orders of magnitude smaller than  $E_{RH}$ without any
fine tuning.

We stress the fact that, although  we will describe the
inflation-dark energy unified scheme choosing  a particular form
of the potential $v(\phi)$, the quantum generation mechanism works
well for a large class of potentials. The only requirements on the
potential $v(\phi)$ are that it should satisfy the slow roll
conditions at high and low energies and that the $\phi$ mass
$m_\phi \equiv \sqrt{v''(\phi)}$ should satisfy the condition $
m_\phi(t_{RH})\gg m_\phi(t_{gen})$, where $t_{RH}$ and $t_{gen}$
are the reheating and $\phi$ generation times. In addition one has
to require that the $\phi$ particles are relativistic at
$t_{gen}$, that implies $m_\phi(t_{gen}) \ll E_{gen}$. In section
\ref{inflationreheating}  we will  discuss the inflation and
reheating scenario and then in section \ref{unification2} we will
consider the inflation and dark energy unified scheme using a
simple example.

\subsection{Inflation and reheating} \label{inflationreheating}

Let us consider an universe that contains the field $\phi$, a
second  scalar field $\varphi$ and the SM particles. We want the
$\phi$ field to inflate the early universe, so we assume that the
potential $v(\phi)$ has at least one flat region corresponding to
inflation. As an example one can consider the potential described
in Appendix \ref{appendix1}. During inflation the $\phi$ field dominates
the universe and slow rolls as long as the slow roll conditions
$|v'(\phi)/v(\phi)| \ll 1$ and $|v''(\phi)/v(\phi)| \ll 1$ are
satisfied and the universe inflates with $H^2 \sim v(\phi)$. After
the end of inflation the $\phi$ field decays into $\varphi$
particles and  reheats the universe. We couple the two scalars $\phi,\vp$ via
the interaction
\begin{equation} \label{generalinteraction}
\textit{L}_{int} = g \, \phi \, f(\varphi)
\end{equation}
where $f(\varphi)$ is a polynomial of $\vp$. We know that the
interaction in eq.(\ref{generalinteraction}) gives a complete
reheating since it involves a single $\phi$ particle decaying into
$\varphi$ particles \cite{kls2}. At the end of inflation the
$\phi$ particles are at rest in the comoving frame (the velocity
is redshifted as $v_i=e^{-\Delta N}v_f$) so $E_\phi = m_\phi$ and
we take $m_\phi \gg m_\varphi$. We take $f(\varphi) = \varphi^3$
and we consider the process $\phi \rightarrow \varphi + \varphi +
\varphi$. For simplicity we assume that all the $\varphi$
particles are produced with the same energy $E_\varphi$ given by
$E_\phi = 3 E_\varphi$ giving a decay rate
\cite{particlesdatabook}
\begin{equation} \label{gammadecay1}
\Gamma_d = \frac{g^2 m_\phi}{ (2 \pi)^3 \, 72}
\end{equation}
Let us remind that a decay process is efficient if $\Gamma_d/H
\gtrsim 1$ \cite{inflationarycosmology}.  The evolution of the
energy density $\rho_\phi$ of the $\phi$ field is given by the
equation
\begin{equation}
\dot\rho_\phi + 3 H (1+\omega) \rho_\phi = - \Gamma_d \, \rho_\phi
\end{equation}
If we consider $\omega$ and $\Gamma_d$ as piecewise constant, the
solution is $ \rho_\phi \sim  {a(t)}^{-3(1+\omega)} \, \, e^{-
\Gamma_d \, t }$  and the $\phi$ energy density vanish
exponentially, that means that $\varphi$ particles are produced
and the energy of the $\phi$ field is transferred to the $\varphi$
field. If one relaxes the hypothesis of constant $\Gamma_d$ one
has $\rho_\phi \sim e^{-\int \Gamma_d \, dt}$ and the condition
for an efficient decay is $\int \Gamma_d \, dt >> 1 $. At the same
time we couple $\varphi$ with SM particles. We take the usual
interaction terms
\begin{equation}\label{intlag2}
L_{int} = h \, \varphi^2 \chi^2,\qquad L_{int} =\sqrt{h} \,
\varphi \,  \bar\psi \psi
\end{equation}
where $\chi$ and $\psi$ are SM scalars and fermions, respectively.
As long as SM particles are relativistic,  valid at
temperatures above $ 1 \, TeV$, the processes $\varphi + \varphi
\leftrightarrow \chi + \chi$ or $\varphi + \varphi \leftrightarrow
\bar\psi + \psi$ given by the interaction in eq.(\ref{intlag2}) have a
transition rate $\Gamma_{RH} = h^2 \, n_\varphi/32 \pi
E_\varphi^2$, and using $n_\varphi =
\zeta(3)\, T_\varphi^3/\pi^2$  one has $\Gamma_{RH} = \zeta(3)
\, h^2 \, E_\varphi/32 \pi^3 \bar{r}^3$.
SM particles are produced at $\Gamma_{RH}/H \equiv E_{RH}/E  > 1 $
where we have defined
\begin{equation}\label{ereheating}
E_{RH} \equiv c_{RH} \, h^2 \, m_{Pl} = 3.2
\left(\fr{h}{10^{-5}}\ri)^2 10^4\,GeV
\end{equation}
and $c_{RH} \equiv \frac{\zeta(3)}{32 \pi^4 \bar{r}} \,
\sqrt{\frac{90 \, \Omega_{rel}}{g_r}}\simeq 10^{-4}$.
If reheating
takes place above $300 \, GeV$ we have $g^{SM}_r = 106.75$ and
$\Omega_r = 1$ so that we can estimate $c_{RH} \simeq 10^{-4}$.
Taking $h\leq 1$, so that $\alpha_h=h^2/4\pi<0.1$, the
maximum energy for reheating would be  $10^{14} \, GeV$.
However the limit for
successful reheating scenario is much lower and it may be as low
as  $10 \, MeV$ \cite{lowinflation,lowinflation2} corresponding  to $h\simeq 10^{-8}$.

Therefore SM particles are produced at energies  $E \leq E_{RH}$
and $\varphi$ is in thermal equilibrium with SM particles with
$T_\varphi = T_\gamma$ where $T_\gamma$ is the photon temperature.
As long as $\vp$ is relativistic $T_\vp \propto T_\gamma$ and if
it remains also in thermal equilibrium we have
\begin{equation}
\Omega_\varphi = \frac{g_\varphi}{g^{SM}_{rel}} \,
\Omega^{SM}_{rel}
\end{equation}
where $g^{SM}_{rel}$ is the number of SM relativistic degrees of
freedom and $g_\varphi = 1$.

\subsection{Inflation-Dark Energy Unification}\label{unification2}

Let us now describe the inflation-dark energy unified picture with
an explicit example. Again we start with two scalar fields coupled
through eq.(\ref{generalinteraction}). To be specific we will
choose $\textit{L}_{int} = g \, \phi \, \varphi^3$ and a scalar
potential $v(\phi)$ that inflates at high and low energies. As
mentioned in the introduction in section \ref{unification} the
choice of the scalar potential is not important, there are a wide
number of possibilities,  and we choose to work with
\begin{equation}\label{potential1}
v(\phi) = \frac{V_I}{2} \left(1-\frac{2}{\pi} \arctan{\frac{
\phi}{f}}\right)
\end{equation}
which has two free parameters $E_I \equiv V_I^{1/4}$ and $f$ with
mass dimension. The potential in eq.(\ref{potential1}) has two
regions $\phi < - ( 2 f/\pi)^{1/3}$ and $\phi > \sqrt{2}$ in which
the slow roll conditions $|v'(\phi)/v(\phi)| \ll 1$ and
$|v''(\phi)/v(\phi)| \ll 1$ are satisfied . The asymptotic
expansions of $v(\phi)$ in these two regions are (see appendix
\ref{appendix1})
\begin{equation}
v(\phi) = \left\{ \begin{array}{ll} V_I\, \left( 1 + \frac{
f}{\pi \phi} \right) \qquad  $for$ \quad \phi < -f\\
\frac{V_I f}{\pi \phi}
\qquad\qquad  \qquad$for$ \quad \phi > f
\end{array}\right.
\end{equation}
Inflation is  associated with  the high energy region $\phi < -( 2
f/\pi)^{1/3}$ with $v(\phi) \simeq V_I = E_I^4$ and dark energy
with the region $\phi > \sqrt{2}$ with $v(\phi) \simeq \frac{v_I
f}{\pi \phi}$. We determine the two free parameters in
eq.(\ref{potential1}) with the constrains coming from inflation
$\frac{\delta\rho}{\rho} = 5.3 \times 10^{-4}$ and from dark
energy density $ \rho_{DE}=3 H_o^2 \Omega_{DE}$. Taking $\phi_o
\simeq \sqrt{2} $ gives $E_I \simeq 100 \, TeV$ and $f = \phi_o \,
v_{DE}/V_I \simeq 10^{-39} \, eV$ where we have reintroduced the
correct mass units in $E_I$ and $f$.

After inflation we want to reheat the universe with the SM
particles. To achieve this we couple $\phi$ and $\varphi$ via the
interaction term $\textit{L}_{int} = g \, \phi \, \varphi^3$ and
$\varphi$ with SM particles as in eq.(\ref{intlag2}). The $\phi$
field decays into $\varphi$ via the process $\phi \rightarrow
\varphi + \varphi + \varphi$ with a decay rate $\Gamma_d =
\frac{g^2  m_\phi}{(2\pi)^3 \, 72}$ given in
eq.(\ref{gammadecay1}). This process starts immediately after
inflation with  $H \simeq E_I^2$. The maximum value of
$\Gamma_d/H$ is when $m_\phi$ is also at its maximum at $\phi \sim
f$ giving $\Gamma_d/H \simeq 10^{36}$. Notice that $\Gamma_d
\simeq 10^{9}$ at its maximum and the lifetime $\tau_\phi =
1/\Gamma_d$ of the $\phi$ particles is such that $\tau / \tau_{pl}
\ll 1$ where $\tau_{pl}$ is the Planck time. Therefore all $\phi$
particles decay and at the end of the reheating and one has
$\Omega_\phi = 0$ and $\Omega_\varphi + \Omega_{SM} = 1$. SM
particles are produced through the interaction with the $\varphi$
field via the interaction given in eq.(\ref{intlag2}) and
described in section \ref{inflationreheating}. This process takes
place for energies $E \leq E_{RH} \equiv c_{RH} \, h^2$ given in
eq.(\ref{ereheating}) with $\Gamma_{RH}/H \geq 1$ and therefore
$\varphi$ and SM particles are in thermal equilibrium giving
$\Omega_\varphi = \frac{g_\varphi}{g_{rel}^{SM}}
\Omega^{SM}_{rel}$ as long as $\varphi$ remains relativistic. As
discussed in section \ref{regeneration} as long as $E_{RH} > E >
E_{gen} \equiv c_{gen} \, g^2$ (c.f eq.(\ref{egen}) ) the universe
contains the SM particles plus $\varphi$ and at energies $E \leq
E_{gen} $ the $\phi$ particles starts to be produced via the
process $\varphi + \varphi \leftrightarrow \varphi + \phi$ with a
decay rate $\Gamma_{gen} = \langle\sigma_{gen}v\rangle \,
n_\varphi$. The inflation-dark energy unification scheme involves
three different quantum processes: The $\phi$ decay into
$\varphi$, the SM particles production and the late time $\phi$
generation. Only the first one (inflaton decay) depends on the
choice of the potential $v(\phi)$, through its mass, the other two
depend only on the size of the couplings $g,h$. These processes
take place   at energies
\begin{equation}\label{ee}
E_I, \;\;  E_{RH} \equiv c_{RH} \, h^2,\;\;E_{gen} \equiv
c_{gen} \, g^2
\end{equation}
where
\begin{equation}
\frac{\Gamma_d}{H} \geq 1, \;\; \frac{\Gamma_{RH}}{H} \equiv
\frac{E_{RH}}{E} \geq 1,\;\; \frac{\Gamma_{gen}}{H} \equiv
\frac{E_{gen}}{E} \geq 1
\end{equation}
and  $c_{RH} \simeq 10^{-4}$ for a reheating temperature above
$300 \, GeV$ and $c_{gen} \simeq 6 \cdot 10^{-4}$ if the $\phi$ is
generated at energies below $1\, MeV$. We stress the fact that the
$\phi$ generation at late times is due to the same interaction
term $\textit{L}_{int} = g \, \phi \, \varphi^3$ that gives the
$\phi$ decay after inflation. The main difference is that at low
energies $E_{gen}$ the mass of $\phi$ is many orders of magnitude
lower than its value at high energies $E_I$, i.e. $m_\phi(t_{RH})
\gg m_\phi(t_{gen})$ where $t_{RH}$ and $t_{gen}$ are the
reheating and $\phi$ generation times respectively. We also
require that $m_\phi(t_{gen}) \ll E_{gen}$ so that $\phi$ is
relativistic at $t_{gen}$ as in section \ref{regeneration}. Notice
that in our model the value of the $\phi$ mass at generation time
is   $m_\phi(t_{gen}) \simeq 10^{-14}\, eV$ (see appendix
\ref{appendix1}) which is much smaller than  $E_{gen} \simeq 1 \,
eV$. However at present times $m_\phi(t_o) \simeq 10^{-33}\, eV$
which is a typical mass for a quintessence (dark energy) field.

After the production of the $\phi$ particles $v(\phi)$ is
generated and the $\phi$ classical evolution will drive the
expansion of the universe as described in section
\ref{regeneration}. Concluding we have shown with a simple example
how the inflation-dark energy unification takes place. Of course
it is possible to choose different scalar potentials $v(\phi)$ or
interaction terms $\textit{L}_{int}$ that gives similar results.

\subsection{ $E_{RH}$ and $E_{gen}$ scales}

The values of $E_{RH}$ and $E_{gen}$ do not depend
on the choice of the potential $v(\phi)$ but only
on the couplings $g,h$. The values of $E_{RH}$ and $E_{gen}$ are fixed in terms of the
couplings $g$ and $h$. In general $g$ and $h$ are free parameters
that should give $E_I \geq  E_{RH} > 10 \, MeV$ and $E_{RH}
\gg  E_{gen} > E_o$ and are given by (see eq.(\ref{ee}))
\begin{equation}
 E_{RH} \equiv c_{RH} \, h^2,\;\;E_{gen} \equiv c_{gen} \, g^2
\end{equation}
and  $c_{RH} \simeq c_{gen} \simeq 10^{-4} $.
An interesting
reduction of parameters is if we take $E_{RH}=\sqrt{E_{gen}} $
(in natural units) which gives $g=h^2$ and
\be
 E_{RH} = \le(\fr{E_{gen}}{1\,eV}\ri)^{1/2} 5\times \;10^4 \;GeV
\ee
Notice that this choice of $g,h$ implies a low reheating temperature
and for $E_{gen}$ as small as $E_o\sim 10^{-3}\,eV$ we have $E_{RH}\geq 1.5\, TeV$.
If we set
\bea\label{couplings}
g &=& h^2 = q \fr{E_I}{m_{pl}}=4 \le(\fr{q}{100 }\ri) \le(\frac{E_I}{100\,TeV}\ri)  10^{-12},
\eea
with $q$ a proportionality constant, we have
\bea\label{egen2}
\;E_{gen} &=& \le(\fr{q}{100 }\ri)^2 \le(\fr{E_I}{100\,TeV}\ri)^{2}
26  \, eV\\
E_{RH} &=&\le(\fr{q}{100 }\ri) \le(\fr{E_I}{100\,TeV}\ri)\; TeV.
\label{erh2}\eea
 The fine structure constants associated to
the two couplings are $\alpha_g \equiv \frac{g^2}{4\pi}, \alpha_h \equiv \frac{h^2}{4\pi}$
and cross sections $\sigma_g,\sigma_{RH}$ are then
\bea
\alpha_g   &=& \le(\fr{q}{100 }\ri)^2 \le(\frac{E_I}{100\,TeV}\ri)^2  10^{-24}\\
\alpha_h &=&  3\le(\fr{q}{100 }\ri) \le(\frac{E_I}{100\,TeV}\ri)  10^{-13}
\eea
The cross section for the generation process are
$\sigma_{gen}=g^2/(32\pi E^2_{gen})$ and $ \sigma_{RH}=h^2/(32\pi E^2_{RH})$
giving
\be\label{sg2}
\sigma_{gen}  =\fr{1}{32\pi c_{gen}^2 g^2}=
\le(\fr{100 }{q}\ri)^2  \le(\fr{100\,TeV}{E_I}\ri)^2  0.1\, pb
\ee
\be\label{srh2}
\sigma_{RH}  =\fr{1}{32\pi c_{RH}^2 h^2}=
 \le(\fr{100 }{q}\ri) \le(\fr{100\,TeV}{E_I}\ri) 10^{-11}\, pb
\ee
We find the relationship between $g$ and $h$ in  eq.(\ref{couplings})
interesting but of course it does not need to hold since in
principle $g,h$ and therefore $E_{gen},E_{RH}$ are independent from
each other and eqs.(\ref{egen2}) and (\ref{erh2}) are equivalent to eqs.(\ref{egen})
and (\ref{ereheating}), respectively.

\section{Phenomenology} \label{phenomenology}

In this section we summarize the main phenomenological
consequences of the dark energy quantum generation. Let us first
discuss the consequences of having the relativistic field
$\varphi$. If $\varphi$ is not contained in the SM then it
represent an extra relativistic degree of freedom. CMB temperature
anisotropies  as well as SDSS and 2dF Large Scale galaxy
clustering, Lyman-$\alpha$  absorption clouds, type Ia Supernovae
luminosity distances and BAO data, can be used to determinate the
value of the effective relativistic degrees of freedom, usually
described in terms of the effective number of neutrinos
$N^{eff}_\nu$. The value of $N^{eff}_\nu$ affects the
matter-radiation equality epoch and thus the ISW effect, so CMB
anisotropies are sensitive to deviations from the standard
cosmological model value of $N^{eff}_\nu \simeq 3.04$. Analysis of
the WMAP data combined with other cosmological data sets  allows
for values of $N^{eff}_\nu$ different from its standard model
value. For example in \cite{melchiorri mena} it is found
$N^{eff}_\nu = 4.6^{+1.6}_{-1.5}$ at $95\%$ c.l., consistently
with other analysis \cite{others}. Moreover BBN is also affected
by $N^{eff}_\nu$, because the number of relativistic degrees of
freedom change the value of the expansion rate of the universe and
than influence the expected primordial Helium abundance. The BBN
bound is $N^{eff}_\nu = 3.1^{+1.4}_{-1.2}$ at $95\%$ c.l.
\cite{BBN,melchiorri mena,BBN2}, that seems to be more stringent
than  bounds coming from  CMB data. Anyhow $N^{eff}_\nu$ can
evolve from the BBN epoch at $T \sim 1 MeV$ to the CMB decoupling
era at $T \sim 1 eV$ \cite{BBN2}, so the different bounds coming
from BBN and CMB are compatible.  In our case the extra
relativistic degree of freedom is represented by the scalar field
$\varphi$  that contributes to $N^{eff}_\nu$ an amount $\delta
N^{eff}_\nu = \frac{4}{7} \left( \frac{T_\varphi}{T_\nu} \right)^4
$. If the $\varphi$  decouples from radiation before
neutrinos,  one has $T_\varphi \leq T_\nu$ and  $\delta
N^{eff}_\nu \leq 4/7 \simeq 0.57$, that is in full agrement with
both BBN and CMB data. If $\varphi$ is coupled with photons one
has $\delta N^{eff}_\nu =  \frac{4}{7} \left(\frac{T_\gamma}{T_\nu}
\right)^4 \simeq 2.2$
that shows some
tension with BBN data but is compatible with CMB data. In any case
the existence of the extra relativistic degree of freedom coming
from the $\varphi$ field is consistent and apparently favored  by
cosmological data.

Another important phenomenological aspect of the dark energy
generation model, concerns the coupling of the $\varphi$ field
with SM particles. In principle  $\varphi$ may be coupled with
electrons, baryons and photons, but strong limits  on the strength
of such couplings comes from astrophysical considerations and
accelerator phyiscs. In fact if coupled with electrons, the
$\varphi$ particles are produced in stars and this fact affects
the evolution of the stars, as studied in \cite{axions}. The
coupling strength between $\varphi$ and electrons should satisfy
the condition $\alpha_{\varphi e e} < 0.5 \times 10^{-26}$.
In the model that we present, SM particles are produced at $E_{RH}
= c_{RH} \, h^2 \, m_{Pl} \geq 10 \, MeV$ that gives $h \geq
10^{-8}$ and a fine structure constant $\alpha = h^2/4\pi \geq
10^{-17}$, therefore the $\varphi$ field cannot be coupled with
electrons. If coupled with baryons, the massless scalar field
$\varphi$ could also generate long range forces
\cite{longrangeforces} with possible observable consequences at
astrophysical and cosmological level. The upper bound coming from
long range force experiments is $\alpha_{\varphi B}  < 10^{-47}$
\cite{G.C} thus the $\varphi$ coupling with baryons should be
excluded. If the $\varphi$ field is coupled with photons via the
axion-like interaction term $\textit{L}_{int} =
\frac{g_{\varphi\gamma\gamma}}{4} \, \varphi \, F_{\mu\nu}
\tilde{F}^{\mu\nu} = - g_{\varphi\gamma\gamma} \, \varphi \,  E
\cdot B$ with $g_{\varphi\gamma\gamma} < 10^{-10} \, GeV^{-1}$
\cite{axions}. The bound on the transition rate is
$\Gamma_{\varphi\gamma\gamma} =
\frac{g_{\varphi\gamma\gamma}m_\varphi^3}{64\pi}  \lesssim 10^{-4}
\, \left(m_\varphi/eV \right)^3 \, eV $. Taking $H = c_H \,
T^2/m_{pl}$  the field  $\varphi$  and  photons are coupled for $T
\lesssim T_{RH}^\gamma\simeq \left( m_\varphi/eV \right)^{3/2} \,
10^2 \, GeV$ when $\Gamma_{\varphi\gamma\gamma}/H \geq  1$.
Therefore if $\varphi$ is coupled with photons, SM particles will
be produced at low reheating temperatures of about $T_{RH}^\gamma
\simeq 10^2 \, GeV$ for $m_\varphi \simeq 1 \, eV$. The coupling
with photons is not ruled out by experimental data.

In conclusion, the $\varphi$ field could be coupled either to
neutrinos, photons, neutral Higgs field or supersymmetric partners
of the SM which are currently searched for at LHC. It is
particularly interesting the case in witch the $\varphi$ is
coupled with neutrinos. In the standard cosmological model the
three neutrinos  free-stream and interact only gravitationally.
Free-streaming  lowers the neutrino perturbations and introduce a
source of anisotropic stress. On the contrary, if one or more
neutrinos are coupled with the scalar field $\varphi$, the
interacting neutrinos behave as a tightly-coupled fluid with
density and velocity perturbations but no anisotropy \cite{cs2}.
This fact also affects the adiabatic sound speed $c_s^2$, that is
equal to $1/3$ for free-streaming neutrinos and it is $0<c_s^2\leq
1/3$ for tightly-coupled neutrinos. Therefore the coupling of the
$\varphi$ field with neutrinos produces many observable
consequences on the  Cosmic Neutrino Background (CNB)
\cite{pierpaoli}.

In alternative to the dark energy generation scheme presented in
section \ref{regeneration}, one can generate the $\phi$ field
without any auxiliary field $\varphi$ but coupling $\phi$ directly
with SM neutrinos via an interaction $L_{int} = \sqrt{g} \, \phi
\bar\nu \nu$. Therefore one or more neutrinos will be
tightly-coupled to the $\phi$ quintessence field and this will
produced observable consequences on the CNB \cite{pierpaoli}.

The $\phi$ sector of the quantum generation model has also an
interesting phenomenology. For example  the inflation dark energy
unified  model presented in section \ref{unification} has low
inflationary and and reheating scale $E_I \simeq 100 \, TeV$ and
$E_{RH} \simeq 1TeV$ for $h \simeq 10^{-6}$. This  low reheating
energy does not affect the reheating efficiency and it also avoids
gravitino overabundance problems. From a cosmological point of
view, a low inflationary scale may also affect $N^{eff}_\nu$ as
showed in \cite{lowinflation2}, giving one more possible test of
the unified model. Moreover interesting effects may be observed in
accelerators, as for example at LHC, with energy scales not so far
from the inflationary energy. Then, as discussed above, a rich
phenomenology exists, that  may be used to constrain or falsify
cosmological models that make use of the dark energy generation
mechanism.

\section{conclusions} \label{conclusions}

We will now present a summary and conclusions of our work. One of
the main  goals of this paper was to understand why the dark
energy is manifested at such a late time. To achieve this we have
presented a novel idea,  the quantum generation of dark energy,
giving  a new interpretation of the late time emergence of DE in
terms of a late time quantum production of the quintessence $\phi$
particles. We take a  $2\leftrightarrow 2$ quantum  process
between $\phi$ and a relativistic  $\vp$ particles. The scale
where the $\phi$ field is generated  is dynamically determined by
the condition $\Gamma/H=E_{gen}/E\geq 1$ giving an energy scale
$E\leq E_{gen} $ with $E_{gen} = c_{gen} \,  g^2 \, m_{pl}$ and
$c_{gen} \simeq 6 \cdot 10^{-4}$. Therefore the smallness of
$E_{gen}$ is due to a small coupling $g$ and for $g\simeq
10^{-12}$ gives a $E_{gen } \simeq 1 \, eV$ and a cross section
$\sigma_{gen}\simeq 1 pb$. The acceleration of the universe is
then due to the classical evolution of $\phi$ and determined by
the scalar potential $v(\phi)$. We have described in section
\ref{regeneration}  a universe that initially contains no $\phi$
particles,  i.e. $n_\phi=\Omega_\phi=\dot\phi=v(\phi)=0$, and once
the relativistic particles $\phi$ are produced they become  a
source term for the generation of the scalar potential $v(\phi)$.
Once $v(\phi)$ has been produced the classical equation of motion
gives the evolution of $\phi$.

We show  in section \ref{unification} that it is possible to unify
inflation and dark energy using the same quintessence field
$\phi$. To achieve the unification we required  that the potential
$v(\phi)$ has two flat regions, at high energy for inflation and
low energy for dark energy. In this scenario, after inflation the
field $\phi$ decays completely and reheats the universe with
standard model particles. The universe expands then in a
decelerating  way dominated first by radiation and later by
matter. At low energies the same interaction term that gives rise
to the inflaton decay accounts for the re-generation of the $\phi$
field giving rise to dark energy. An important difference in the
quantum process between $\phi$ and $\vp$ at high and low energies
is the value of transition rate due to the size of the $\phi$
mass,  $m^2_\phi(E_I)\gg m^2_\phi(E_o)$.

We presented  in section \ref{unification}  a simple example on
how the inflation-dark energy unification can be implemented. We
used a potential $v=E_I^4(1-\arctan[\phi/f])$ which is flat at
high and low energies. The  two parameters $E_I, f$ are determined
by the density perturbations  $\delta\rho/\rho$ and the value of
$v_o$ at present time giving $E_I=100\,TeV, f=10^{-39} eV$. The
coupling $g$  between $\phi$ and $\vp$ and the coupling $h$
between $\vp$ and the SM particles are free parameters but can be
taken as $g=h^2= q \, E_I/m_{pl}=(q/100)(E_I/100\,TeV)10^{-12}$
giving a reheating energy $E_{RH}=(q/100)(E_I/(00\,TeV)\, TeV$ and
a generation energy  $ E_{gen}= (q/100)^2(E_I/100\,TeV)^2 26  \,
eV$. The cross section between $\phi$ and $\vp$ is $\sigma_{gen} =
g^2/32 \pi E_{gen}^2 \simeq  pb$ quite close to cross section of
WIMP dark matter with nucleons $\sigma_{w} \simeq
 pb$ at decoupling. By fixing $g=h^2=q\,E_I$ we have determined the
coupling, which set the scales of reheating and $\phi$
re-generation, in terms of the inflation scale $E_I$ and we can
reduced the number of parameters. Of course this is not the only
possible choice of $g$ and $h$.\\
To conclude, we have presented a general framework to produce the
fundamental quintessence field $\phi$ dynamically at low energies.
The energy scale is fixed by the strength of the coupling and this
offers a new interpretation of the cosmological coincidence
problem:   dark energy domination starts at such small energies
because of  the size of the coupling constant $g$. Finally, our
approach  allows for an easy implementation of inflation and dark
energy unification with the standard long periods of
radiation/matter domination.

\appendix

\section{Equations of motion} \label{appendix2}

Here we derive the system of differential
eqs.(\ref{equation1})-(\ref{equation4}) that rules the $\phi$
generation. Let us consider a FRW universe containing the $\phi$
field coupled with a second scalar field $\varphi$ and let us
write down the equations of motion of the   $\phi$ and $\varphi$
fields. In what follows we assume that both the $\phi$ and
$\varphi$ particles are relativistic. The $\phi(t,x)$ field can be
divided into a classical background configuration $\phi_c(t)$ plus
a perturbation $\delta\phi(t,x)$ corresponding to the quantum
configuration of the $\phi$ field ($\phi$ particles), in such a
way that we can write $\phi(t,x) = \phi_c(t) + \delta\phi(t,x)$.
We choose $\phi(t,x)$ and $\delta\phi(t,x)$ as independent
variables, stressing the fact that when $\delta\phi(t,x)
\rightarrow 0$ one has $\phi(t,x) = \phi_c(t)$. The dynamic of the
$\phi$ field is then deduced from its lagrangian density
\begin{equation}
\textit{L} = \frac{1}{2}\partial_\mu\phi \partial^\mu \phi +
\frac{1}{2}\partial_\mu\varphi \partial^\mu \varphi -
V_T(\phi,\varphi)
\end{equation}
where $V_T = v(\phi) +B(\varphi) + v_{int}(\phi,\varphi)$,
$v(\phi)$ and $B(\varphi)$ are the classical potentials of the
$\phi$ and $\varphi$ fields respectively and  $
v_{int}(\phi,\varphi) = -\textit{L}_{int}(\phi,\varphi)$ where
$\textit{L}_{int}(\phi,\varphi)$ is the interaction lagrangian of
the scalar fields $\phi$ and $\varphi$. Using $\nabla \phi(t,x) =
\nabla \delta\phi(t,x)$  one has
\begin{equation}\label{classicalequation1}
\ddot{\phi} + 3 H \dot\phi - \frac{ \nabla^2\delta\phi}{a^2} +
\partial_\phi V_T(\phi,\varphi) = 0
\end{equation}
The perturbation $\delta\phi$ evolves as a scalar field with mass
$m^2_{\phi} = \partial^2_\phi v(\phi)$. The corresponding quantum
operator $\delta\hat{\phi}$ has the following expression
\be
\delta\hat{\phi} = \int \fr{d^3k}{(2\pi)^3\sqrt{2E_k}} \left[ a_k \,f(t)
\,e^{-i\vec{k}\vec{x}} + a_k^\dag \,f^*(t)
\,e^{i\vec{k}\vec{x}}\right]
\ee
where $k$ and
$E_k$ are respectively the wave number and the energy of the
$\phi$ particles with $E_k^2 = |\vec{k}|^2/a(t)^2 + m^2_\phi$
\cite{G},\cite{mandleshaw}. The physical momentum is $\vec{p} =
\vec{k}/a(t)$ and the $\phi$ particles are relativistic so $E_k =
|\vec{k}|/a(t) = |\vec{p}|$. For simplicity we assume that all the
$\phi$ and $\varphi$ particles have the same energy $E_\phi$ and
$E_\varphi$ respectively. For example this assumption is valid if
the quantum particles are thermalized. In this case the phase
space distributions are the Bose-Einstein distributions $f_\phi(E)
= 1/ (e^{E/T_\phi} -1)$  and $f_\varphi(E) = 1/ (e^{E/T_\varphi}
-1)$ and one can take the energies of the $\phi$ and $\varphi$
particles as the mean values $E_\phi = \bar{r} \, T_\phi$ and
$E_\varphi = \bar{r} \, T_\varphi$ with $\bar{r} \simeq 2.7$.
Moreover, in the case that the $\phi$ and $\varphi$ fields are in
thermal equilibrium one has $E_\phi = E_\varphi$. Therefore we
estimate $\delta\phi^2$ as the average of $\delta\hat\phi$ on the
quantum state $|N,E_\phi>$ containing $N_{\phi}$ particles with
energy $E_{\phi}$, i.e. $ <N,E_\phi|: \delta\hat{\phi}^2
:|N,E_\phi> = \frac{n_{\phi}}{E_{\phi}} $, where the $::$  stands
for normal ordering of creation and destruction operators. We have then
$\delta\phi^2 = n_{\phi}/E_{\phi}$, where $n_\phi$ is the density
number of $\phi$ particles. Thus we can write
\begin{equation}\label{nablaphi}
- \frac{\nabla^2 \delta\phi}{a^2} = |\vec{p}|^2 \delta\phi =
E_{\phi}^{3/2} \sqrt{n_{\phi}}
\end{equation}
where we have used $E_k^2 = |\vec{p}|^2 = |\vec{k}|^2/a^2$ for
relativistic particles. Substituting this expression in
eq.(\ref{classicalequation1}) we have
\begin{equation}\label{phievolution}
\ddot{\phi} + 3 H \dot\phi + E_{\phi}^{3/2} \sqrt{n_{\phi}} +
\partial_\phi v(\phi) + \partial_\phi v_{int}(\phi,\varphi) = 0.
\end{equation}
Repeating the same considerations for the $\varphi$ field one has
\begin{equation}\label{varphievolution}
\ddot{\varphi} + 3 H \dot\varphi + E_{\varphi}^{3/2}
\sqrt{n_{\varphi}} + \partial_\varphi B(\varphi) +
\partial_\varphi v_{int}(\phi,\varphi) = 0.
\end{equation}
The dynamics of the quantum particles is governed by the Boltzmann
equations. If $f_\phi(E,t)$ is the phase space distribution of the
$\phi$ particles, one has $n_\phi = \int f_\phi(E,t)
\frac{d^3p}{(2\pi)^3}$, $\rho_\phi = \int E \, f_\phi(E,t)
\frac{d^3p}{(2\pi)^3}$ and $p_\phi = \int \frac{|\vec{p}|^2}{3 E}
f_\phi(E,t) \frac{d^3p}{(2\pi)^3}$. In the same way one has
$n_\varphi = \int f_\varphi(E,t) \frac{d^3p}{(2\pi)^3}$,
$\rho_\varphi = \int E \, f_\varphi(E,t) \frac{d^3p}{(2\pi)^3}$
and $p_\varphi = \int \frac{|\vec{p}|^2}{3 E} f_\varphi(E,t)
\frac{d^3p}{(2\pi)^3}$ where $f_\varphi(E,t)$ is the phase space
distribution of the $\phi$ particles. The evolution of the phase
space density $f_\phi(E,t)$ is governed by the Boltzmann equation
$\hat L[f_\phi] = \hat C[f_\phi]$, where $\hat L$ is the Liouville
operator and  in a FRM metric is $\hat L[f_\phi(E,t)] = E \,
\partial_t f_\phi(E,t) - H |\vec{p}|^2 \,
\partial_{E} f_\phi(E,t)$ and $\hat C$ is the collision
operator (see \cite{kolbturner}). If the $\phi$ particles are
relativistic one has
\begin{equation}\label{Boltzmanequation1}
\dot{n}_\phi + 3 H n_\phi = \int \hat C[f_\phi(E,t)] \,
\frac{d^3p}{(2\pi)^3 \, E} +  A
\end{equation}
where for a process $ a_1 +a_2+...+a_n \leftrightarrow
b_1+b_2+...+b_l$, with $a_n (b_l)$ initial (final) particles, one
has
\bea \label{collisionoperator}&\int& C[f_\phi(E,t)] \,
\frac{d^3p}{(2\pi)^3 \, E} =
-\int d\Pi_{a_1}...d\Pi_{a_n} d\Pi_{b_1}...d\Pi_{b_l} \nonumber\\
&& \times (2\pi)^4\,|M_{ab}|^2 \delta^4(\Sigma_i^n
P_{a_i}-\Sigma_j^l  P_{b_j})\\
&& \times \left[f_{a_1}(E,t)...f_{a_n}(E,t)- f_{b_1}(E,t)... f_{b_l}(E,t) \right]
\nonumber\eea
with $d\Pi \equiv g \, d^3p/(2\pi)^3 2E$, $g$ are
the internal degrees of freedom and $|M_{ab}|^2$ the transition
scattering matrix of the process.
Eq.(\ref{collisionoperator}) is valid in absence of Bose
condensation of Fermi degeneracy when $1 \pm f_i(E,t) \simeq 1$
\cite{kolbturner}. In the same way one has
\begin{equation}\label{Boltzmanequation2}
\dot{n}_\varphi + 3 H n_\varphi = \int \hat C[f_\varphi(E,t)] \,
\frac{d^3p}{(2\pi)^3 \, E} +  Q
\end{equation}
The  terms $A$ and $Q$ introduced in eqs.(\ref{Boltzmanequation1})
and (\ref{Boltzmanequation2}) are necessary for the energy
conservation as we will discuss  below  eqs.(\ref{Acond}) and
(\ref{Bcond}).

Let us consider the quadratic interactions $\textit{L}_{int} = g
\, \phi^2 \, \varphi^2$ or $\textit{L}_{int} = g \, \phi \,
\varphi^3$. In that case the $\phi$ field is generated via the
$2\leftrightarrow 2$ processes $\varphi + \varphi \leftrightarrow
\phi + \phi$ or $\varphi + \varphi \leftrightarrow \varphi + \phi$
respectively. For simplicity we assume that the phase space
distribution $f_\varphi(E,t)$ of the $\varphi$ particles is piked
around the mean energy $E_\varphi$ of the  $\varphi$ particles. Of
course this is true in the case of thermalized particles.
Therefore we take all the $\varphi$ particles with the same energy
$E_\varphi$ and one has $ \int f_\varphi(E,t) \,d\Pi_\varphi =
\int f_\varphi(E,t) \, \frac{d^3p_\varphi}{2 E_\varphi (2\pi)^3}
\simeq \frac{1}{2 E_\varphi} \int f_\varphi(E,t)
\frac{d^3p_\varphi}{ (2\pi)^3} = \frac{n_\varphi}{2 E_\varphi}$.
Moreover, from energy conservation it follows that the $\phi$
particles are produced with the same energy of the $\varphi$
particles and we take $E_\phi = E_\varphi = E$, that  is also
valid when the $\phi$ and $\varphi$ fields thermalize. Therefore
the energy distribution of the $\phi$ particles is piked around
the mean energy $E_\phi$ and one also has $\begin{array} {ll} \int
f_\phi(E,t) \,d\Pi_\phi = \frac{n_\phi}{2 E_\phi}
\end{array}$.
The transition rates for the considered processes are
$\Gamma_{\varphi \varphi \rightarrow \phi \phi} = \Gamma_{\varphi
\varphi \rightarrow \phi \varphi} = \Gamma_{\phi \varphi
\rightarrow \varphi \varphi} =  \langle\sigma_{gen}v\rangle \,
n_\varphi\equiv \Gamma_{gen}$ and $\Gamma_{\phi \phi \rightarrow
\varphi \varphi} = \langle\sigma_{gen}v\rangle \, n_\phi$, where
$\sigma_{gen}= g^2/32 \pi E^2$ is the cross section for a $2
\leftrightarrow 2$ relativistic particle process and $v$ is the
relative velocity \ci{G}. Considering the process  $\varphi
\varphi \leftrightarrow \varphi \phi$ one has
\begin{equation}\label{Gamma1}
\begin{array}{ll}
\int C[f_\phi(E,t)] \, \frac{d^3p}{(2\pi)^3 \, E} = - \int
C[f_\varphi(E,t)] \, \frac{d^3p}{(2\pi)^3 \, E} =\\
= \Gamma_{\varphi \varphi \rightarrow \phi \varphi}\,n_\varphi^2
- \Gamma_{\phi \varphi \rightarrow \varphi \varphi} n_\varphi \,
n_\phi = \langle\sigma_{gen}v\rangle \, n_\varphi (n_\varphi -
n_\phi)
\end{array}\end{equation}
and for the process  $\varphi \varphi \leftrightarrow \phi \phi$
one has
\begin{equation}\label{Gamma2}
\begin{array}{ll}
\int C[f_\phi(E,t)] \, \frac{d^3p}{(2\pi)^3 \, E} = - \int
C[f_\varphi(E,t)] \, \frac{d^3p}{(2\pi)^3 \, E} =\\
= \Gamma_{\varphi \varphi \rightarrow \phi \phi}\,n_\varphi^2 -
\Gamma_{\phi \phi \rightarrow \varphi \varphi} \, n_\phi^2 =
\langle\sigma_{gen}v\rangle  (n_\varphi^2 - n_\phi^2) = \\
= \langle\sigma_{gen}v\rangle (n_\varphi + n_\phi) (n_\varphi -
n_\phi)
\end{array}\end{equation}
We can write eqs.(\ref{Gamma1}) and (\ref{Gamma2}) in a compact
form as
\bea
\int C[f_\phi(E,t)] \, \frac{d^3p}{(2\pi)^3 \, E} &=& - \int
C[f_\varphi(E,t)] \, \frac{d^3p}{(2\pi)^3 \, E}= \nonumber \\
&=& \, \tilde{\Gamma} \left( n_\varphi - n_\phi \right)
\eea
where we have defined  $\tilde{\Gamma} \equiv
\langle\sigma_{gen}v\rangle n_\varphi$ for the process $\varphi
\varphi \leftrightarrow \varphi \phi$ and $\tilde{\Gamma} \equiv
\langle\sigma_{gen}v\rangle (n_\varphi+n_\phi)$ for the process
$\varphi \varphi \leftrightarrow \phi \phi$. Note that
$\tilde{\Gamma}$ is not necessarily a transition rate, but it
accounts for the whole contribution of the two processes
$\varphi \varphi \rightarrow \varphi \phi$ and $\varphi \phi
\rightarrow \varphi \varphi$ in one case and $\varphi \varphi
\rightarrow \phi \phi$ and $\phi \phi \rightarrow \varphi \varphi$
in the other case.

Therefore eqs.(\ref{Boltzmanequation1}) and
(\ref{Boltzmanequation2}) now read
\bea\label{Boltzmanequation3}
\dot{n}_\phi + 3 H n_\phi &=& \tilde{\Gamma} \left( n_\varphi -n_\phi\right) +  A\\
\dot{n}_\varphi + 3 H n_\varphi &=& - \tilde{\Gamma} \left(n_\varphi - n_\phi \right) +  Q.
\label{Boltzmanequation4}
\eea
The
energy density and pressure of the system are
\bea\label{totalenergydensity}
\rho_T &=& \frac{\dot\phi^2}{2} + \frac{\dot\varphi^2}{2} +
V_T(\phi,\varphi) +  \frac{E_\phi \, n_\phi}{2} +\frac{E_\varphi
\, n_\varphi}{2}\\
p_T &=& \frac{\dot\phi^2}{2}    + \frac{\dot\varphi^2}{2} -
V_T(\phi,\varphi) +  \frac{E_\phi \, n_\phi}{6} +\frac{E_\varphi
\, n_\varphi}{6}
\label{totalpressure}\eea
with $V_T(\phi,\varphi)=v(\phi)+B(\vp)+v_{int(\phi,\vp)}$.
Note that the terms proportional to the number densities in
eqs.(\ref{totalenergydensity}) and (\ref{totalpressure}) comes
from the terms $|\nabla \delta\phi|^2/a(t)^2 = E_\phi^2
\delta\phi^2 = E_\phi n_\phi$ and $|\nabla \delta\varphi|^2/a(t)^2
= E_\varphi^2 \delta\varphi^2 = E_\varphi n_\varphi$

Therefore eqs.(\ref{totalenergydensity}) and
(\ref{totalpressure}), together with
eqs.(\ref{phievolution}),(\ref{varphievolution}),(\ref{Boltzmanequation3})
and (\ref{Boltzmanequation4}) give the energy conservation in the
form
\bea\label{energyconservation} \dot \rho_T &+& 3 H \left(
\rho_T + p_T  \right) = - E_\phi^{3/2} \sqrt{n_\phi} \,
\dot\phi + \frac{A\, E_\phi}{2} - \nonumber  \\&-& E_\varphi^{3/2}
\sqrt{n_\varphi} \, \dot\varphi + \frac{Q \, E_\varphi}{2} = 0
\eea
Note that eq.(\ref{energyconservation}) should be
valid in the case in which only one of the scalar fields $\phi$
and $\varphi$ exists. For example, if the only $\phi$ field exists
one has $v_{int}(\phi,\varphi) =0 $, $n_\varphi = 0$ and $Q = 0$
and  form eq.(\ref{energyconservation}) one has
\begin{equation}\label{Acond}
A = 2  \sqrt{E_\phi \, n_\phi} \, \dot\phi
\end{equation}
and in the case in which the only $\varphi$ field exists
one obtains
\begin{equation}\label{Bcond}
Q = 2 \sqrt{E_\varphi \, n_\varphi} \, \dot\varphi.
\end{equation}
Then
eqs.(\ref{phievolution}),(\ref{varphievolution}),(\ref{Boltzmanequation3})
and (\ref{Boltzmanequation4}) now read
\begin{equation}\label{systemregeneration}
\begin{array}{ll}
\ddot{\phi} + 3 H \dot\phi + E_{\phi}^{3/2} \sqrt{n_{\phi}} +
\partial_\phi v(\phi) +
\partial_\phi v_{int}(\phi,\varphi) = 0\\
\ddot{\varphi} + 3 H \dot\varphi + E_{\varphi}^{3/2}
\sqrt{n_{\varphi}} + \partial_\varphi B(\varphi) +
\partial_\varphi v_{int}(\phi,\varphi) = 0\\
\dot{n}_\phi + 3 H n_\phi =  \, \tilde{\Gamma} \left( n_\varphi -
n_\phi \right) +  2 \sqrt{E_\phi \, n_{\phi}}  \, \dot\phi
\\
\dot{n}_\varphi + 3 H n_\varphi = -  \, \tilde{\Gamma} \left(
n_\varphi - n_\phi \right) +  2 \sqrt{E_\varphi \, n_{\varphi}}
\dot\varphi
\end{array}\end{equation}
The system in eqs.(\ref{systemregeneration}) describe the dynamics
of two coupled relativistic scalar fields with the same energy
$E_\phi = E_\varphi = E$. Note that this system includes the
quantum interaction between the quantum  $\phi$ and $\varphi$
particles trough the term $\tilde{\Gamma} \left( n_\varphi -
n_\phi \right) $ in the last two equations in the system
(\ref{systemregeneration}). Moreover the density numbers $n_\phi$
and $n_\varphi$ generates a source term for the corresponding
fields $\phi$ and $\varphi$ in first two equations of the system
(\ref{systemregeneration}) and this source term is responsible of
the generation of the classical potential $v(\phi)$ during the
$\phi$ generation. It is useful to divide the energy density
$\rho_T$ in two terms $\rho_T = \rho_\phi + \rho_\varphi$ with
$\rho_\phi = \rho_{1\phi} + \rho_{2\phi}$,
\begin{equation}
 \rho_{1\phi} =
\frac{\dot\phi^2}{2} + v(\phi) \qquad
\rho_{2\phi} = \frac{E_\phi \, n_\phi}{2}
\end{equation}
and $\rho_\varphi = \rho_{1\varphi} + \rho_{2\varphi}$,
\begin{equation}
\rho_{1\varphi} = \frac{\dot\varphi^2}{2} +  v_{int}(\phi,\varphi)+ B(\varphi),\;\;
\rho_{2\varphi} = \frac{E_\varphi \, n_\varphi}{2}.
\end{equation}
In the same way we can write the pressure of the system as $p_T =
p_\phi + p_\varphi$ with $p_\phi = p_{1\phi} + p_{2\phi}$,
\begin{equation}
p_{1\phi} = \frac{\dot\phi^2}{2} - v(\phi),\qquad p_{2\phi} = \frac{E_\phi \,
n_\phi}{6}
\end{equation}
and $p_\varphi = p_{1\varphi} + p_{2\varphi}$,
\begin{equation}
p_{1\varphi} = \frac{\dot\varphi^2}{2} -
v_{int}(\phi,\varphi)- B(\varphi),\;\;
p_{2\varphi} = \frac{E_\varphi \, n_\varphi}{6}.
\end{equation}
It is easy to cheek that these quantities verifies the following
evolutionary equations
\begin{equation}\label{energyconservationphi}
\begin{array}{ll}
\dot \rho_{1\phi} + 3 H ( \rho_{1\phi} + p_{1\phi} ) = -\dot\phi
\, \partial_\phi v_{int} -  \dot\phi E_\phi^{3/2} \,
\sqrt{n_\phi}  \\
\dot \rho_{2\phi} + 3 H ( \rho_{2\phi}  + p_{2\phi}  ) =
\frac{1}{2} \, E_\phi \tilde{\Gamma} ( n_\varphi -
n_\phi ) + \dot\phi E_\phi^{3/2}  \sqrt{n_\phi} \\
\dot \rho_{\phi} + 3 H ( \rho_{\phi}  + p_{\phi}) = \frac{1}{2}
E_\phi  \tilde{\Gamma}  ( n_\varphi - n_\phi ) - \dot\phi \,
\partial_\phi v_{int}
\end{array}\end{equation}
and
\begin{equation}\label{energyconservationvarphi}
\begin{array}{ll}
\dot \rho_{1\varphi} + 3 H ( \rho_{1\varphi} + p_{1\varphi}) =
\dot\phi \,\partial_\phi v_{int} -  \dot\varphi E_\varphi^{3/2}   \sqrt{n_\varphi}
\\
\dot \rho_{2\varphi} + 3 H  ( \rho_{2\varphi} + p_{2\varphi}) = -
\frac{1}{2}  E_\varphi \tilde{\Gamma}  ( n_\varphi - n_\phi  ) +
\dot\varphi E_\varphi^{3/2}   \sqrt{n_\varphi}
\\
\dot \rho_{\varphi} + 3 H  ( \rho_{\varphi}  + p_{\varphi}) =
-\frac{1}{2}  E_\varphi \, \tilde{\Gamma}  ( n_\varphi - n_\phi  )
+ \dot\phi \,\partial_\phi v_{int}
\end{array}\end{equation}
Notice that the sum of the first two equations of
the systems   in eqs.(\ref{energyconservationphi}) or (\ref{energyconservationvarphi})
gives the last equation, respectively, while the sum of the last equation in
eqs.(\ref{energyconservationphi}) and (\ref{energyconservationvarphi}) gives the total
energy evolution in eq.(\ref{energyconservation})
as it should due to energy-momentum conservation.

\section{The  potential $v(\phi)$}\label{appendix1}

\begin{figure}[tbp]
\begin{center}
\includegraphics*[width=7cm]{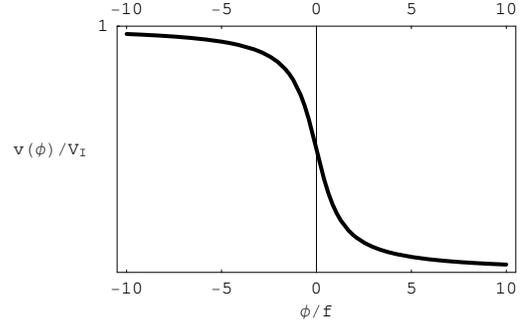} \caption{Plot of the
potential $v(\phi)$.} \label{nombre}
\end{center}
\end{figure}
Let us study in more detail the properties of the potential
\bea\label{potentialv}
 v(\phi) &=& \frac{V_I}{2}
\left(1-\frac{2}{\pi} \arctan{\frac{
\phi}{f}}\right)\\
 v'(\phi) &=& -\frac{V_I}{\pi\, f} \,
\frac{1}{1+(\phi/f)^2}\\
m^2_\phi&\equiv& v''(\phi)  = \frac{V_I}{\pi\, f^3} \,
\frac{\phi}{(1+(\phi/f)^2)^2}
\eea
where $E_I = V_I^{1/4} $ and $f$ are parameters with mass
dimensions. The $\phi$ mass is maximum at $|\phi| \simeq f$ with
$m^2_\phi \simeq V_I/f^2$ while $v'$ is always negative.  The
asymptotic expansion of the potential in eq.(\ref{potentialv}) for
$|\phi/f|\gg 1$ is
\begin{equation} \label{asympt2}
v(\phi) \simeq \left\{
\begin{array}{ll}
V_I\, \left( 1 + \frac{ f}{\pi \phi} \right) \qquad $for$ \quad
\phi < - f\\
\frac{V_I f}{\pi \phi} \qquad\qquad  \qquad $for$ \quad \phi > f
\end{array}
\right.
\end{equation}
One can easily cheek that the slow roll conditions
$|v'(\phi)/v(\phi)| \ll 1$ and $|v''(\phi)/v(\phi)| \ll 1$ are
satisfied in the two regions $\phi < - ( 2 f/\pi)^{1/3}$ and $\phi
> \sqrt{2}$. Therefore the region $\phi < - ( 2 f/\pi)^{1/3}$
is associated with inflation at energies $E = \sqrt{\rho_\phi}
\simeq E_I$. Dark energy is associated to the region $\phi >
\sqrt{2}$ at energy $E  = \sqrt{\rho_\phi} \simeq E_I
(f/\phi_o)^{1/4} \simeq E_{DE} \simeq 3 \times 10^{-3} eV$ where
we have chosen $\phi_o \simeq \sqrt{2}$ as the present time value
of $\phi$. We can fix the parameters $E_I,f$ by imposing
$\delta\rho/\rho = 5.3 \times 10^{-4}$ and from dark energy
density $ \rho_{DE}=3 H_o^2 \Omega_{DE}$, this gives $E_I \simeq
100 \, TeV$ and $f  \simeq 10^{-39} \, eV$, which gives a present
time mass $m_\phi(t_o) \simeq 10^{-33} \, eV$ which is the
standard value for quintessence field.

We can express the value of $\phi$ at the generation time, i.e.
$\phi_{gen}$, in terms of $\phi_o$ as $\phi_{gen} =
\frac{v(\phi_{DE})}{v(\phi_{gen})}\, \phi_o =
(\frac{E_{DE}}{E_{gen}})^4 \, \phi_o$ and if one choose $E_{gen}
\simeq 1 \, eV$ one has $\phi_{gen} \simeq 10^{-12}\, \phi_o$.
Therefore we can also compare the value of the mass $m_\phi$ at
the reheating, generation and present times. At reheating one has
 $\phi \simeq f$ and $m_\phi(t_{RH}) \simeq \sqrt{V_I
/f^2}$, at generation time one has $\phi_{gen} \gg f$ and
$m_\phi(t_{gen}) \simeq \sqrt{\frac{V_I f}{\pi \phi^3_{gen}}}$ and
at present time time one has $\phi_{o} \gg f$ and $m_\phi(t_{o})
\simeq \sqrt{\frac{V_I f}{\pi \phi^3_{o}}}$. One has
$m_\phi(t_{RH})/m_\phi(t_{gen}) \simeq \sqrt{\phi_{gen}/f} \gg 1
$, therefore there are many orders of magnitudes of difference
between the values of $m_\phi$ at the reheating and generation
times, and this is why the reheating is obtained via the decay
process $\phi \rightarrow \varphi + \varphi + \varphi$ of massive
$\phi$ particles into relativistic $\varphi$ particles, and the
$\phi$ is generated at late times via a $2 \leftrightarrow 2$
process between relativistic particles.

We can also estimate the number of e-folds during inflation as $N
= \ln\frac{a_f}{a_i} = - \int_{\phi_i}^{\phi_f}
\frac{v\left(\phi\right)}{ v'\left(\phi\right)} d\phi $. During
inflation one has $\phi \leq - \left(2f/\pi\right)^{1/3} < f$,
therefore we can use the second asymptotic expansion in
eq.(\ref{asympt2}) to write
\begin{equation}
N = - \int_{\phi_i}^{\phi_f} \frac{\pi\phi^2}{f} =
\frac{\pi}{3f} \left(\phi_f^3-\phi_i^3\right)
\end{equation}
Therefore if one require a minimum number $N_m$ of e-folds during
inflation, inflation must start at $\phi_i \leq \left(\phi_f^3- f
N_m/\pi\right)^{1/3} \simeq - \left(f/\pi \right)^{1/3} \,
\left(N_m +2\right)^{1/3}$ for $\phi_f \simeq -
\left(2f/\pi\right)^{1/3}$. Note that a reasonable number of
e-folds $N_m \simeq 50-100$ is easily achieved in the
interval $ \phi \in [- \left(f/\pi \right)^{1/3} \, \left(N_m
+2\right)^{1/3} , - \left(2f/\pi \right)^{1/3} ]$ of width $\Delta
\phi \sim \phi \sim f^{1/3}$.

\end{document}